\documentclass[iop]{emulateapj}
\shorttitle{Inclination Excitation due to Stellar Flybys}
\shortauthors{Moore et al.}

\usepackage{color}

\usepackage{amsmath,graphicx,natbib}

\begin{document}

\title{Inclination Excitation of Solar System Debris Disk due to Stellar Flybys}

\author{Nathaniel W. H. Moore\altaffilmark{1}, Gongjie Li\altaffilmark{1}, 
Fred C. Adams\altaffilmark{2,3}}
\affil{$^1$ Center for Relativistic Astrophysics, School of Physics, Georgia Institute of Technology, Atlanta, GA 30332, USA}
\affil{$^2$ Physics Department, University of Michigan, Ann Arbor, MI 48109, USA}
\affil{$^3$ Astronomy Department, University of Michigan, Ann Arbor, MI 48109, USA}

\email{natemo13@gatech.edu}

\begin{abstract}
Most stars form in clusters where relatively close encounters with other stars are common and can leave imprints on the orbital architecture of planetary systems. In this paper, we investigate the inclination excitation of debris disk particles due to such stellar encounters. We derive an analytical expression that describes inclination excitation in the hierarchical limit where the stellar flyby is distant. We then obtain numerical results for the corresponding particle inclination distribution in the non-hierarchical regime using a large ensemble of N-body simulations. For encounters with expected parameters, we find that the bulk inclination of the disk particles remains low. However, a distinct high inclination population is produced by prograde stellar encounters for particles with final pericenter distances above $50$AU. The maximum extent $i_t$ of the inclination distribution scales with the inclination of the encounter $\sin(i_s)$ for massive star flybys with low incoming velocity. The inclination distribution of observed trans-Neptunian objects places constraints on the dynamical history of our Solar System. For example, these results imply an upper limit on product of the number density $n$ of the solar birth cluster and the Sun's residence time $\tau$ of the form $n\tau\lesssim8\times10^4$ Myr pc$^{-3}$. Stronger constraints can be derived with future observational surveys of the outer Solar System.
\end{abstract}

\section{Introduction} \label{sec:intro}

Objects found in the outskirts of the Solar System, beyond the orbit of Neptune, display interesting orbital properties, and the formation of the observed Solar System architecture requires a theoretical explanation. In particular, a growing number Trans-Neptunian objects (minor or dwarf planets that orbit the Sun beyond Neptune) have been discovered and some of them are found on highly inclined orbits \cite[e.g.,][]{brown_discovery_2004,Chen13, trujillo_sedna-like_2014, Chen16, Shankman17, becker_discovery_2018}. Some of these objects, like Sedna, orbit with a pericenter distance of greater than $50$AU. With their distant orbits, these objects avoid close encounters with Neptune, so that additional dynamical mechanisms are required to account for their eccentric and inclined orbits. 

Several mechanisms have been proposed to explain the origin of these Sedna-like objects with high inclination. For instance, an additional (as yet undetected) new planet on a wide and eccentric orbit can influence the orbits of Sedna-like objects and enhance their inclinations. Such a model is able to explain other features of the outer Solar System, such as the alignment of the pericenters of TNOs \cite[e.g.,][]{trujillo_sedna-like_2014,batygin_evidence_2016,Li18,plan9review}. Alternately, these orbits can also be produced and/or captured from a passing planetary or stellar system. \cite[e.g.,][]{kenyon_stellar_2004, morbidelli_scenarios_2004, Brasser06, jilkova_how_2015}. The Nice model, in which the inner giant planets formed in a more compact configuration and subsequently migrated outwards, has also been used to explain some architectural features of the outer Solar System. During the migration of the inner planets, disc objects in the outer Solar System could have been scattered to higher eccentricity and inclination values \citep[]{fernandez_ip84,hahn_malhotra99,thommes_etal99,gomes03,tsiganis_origin_2005,morbidelli_etal05,levison_etal08,nesvorny15}. However, the migration of the planets alone is insufficient to explain the inclined orbits of detached objects with pericenter distances greater than $50$AU as these planets are far enough away to avoid scattering. 
Moreover, \citet{Siraj20} recently proposed that the Solar System could once had a temporary stellar companion in the birth cluster, and this could also enhance the inclination of the detached objects.

Given that most stars form within groups and clusters \citep{carpenter_cluster_10, lada_embedded_2003, porras_catalog_2003}, the Solar System is likely to have experienced stellar encounters in the past. In particular, it has been argued that our Sun is likely to have formed in a region of relatively high stellar density, based on  meteoritic enrichment in short-lived radiogenic isotopes as well as the orbital structure of the outer Solar System \citep[see the review of][]{adams_birth_2010}. As a result, close encounters of the Solar System with other stars were much more common than they are today. Moreover, the encounters took place at lower speeds, $\sim1$ km/s, allowing for more disruptive interactions. Such encounters in the birth cluster can change the orbits of the planets and planetesimals orbiting around the stars thereby leaving imprints on planetary architectural features, including disk truncation \citep{winter_protoplanetary_2018,winter_protoplanetary_2018-1}. Recently, many studies have shown that stellar flybys can perturb the orbits of planets/objects with wide orbits and leave unique signatures in the planetary architecture \citep[e.g.,][]{Cai19, koncluster2020, Li20, Wang20}. For our own Solar System, \cite{pfalzner_outer_2018} used N-body simulations to show that stellar encounters from a system's birth cluster can produce TNOs with orbital features similar to those observed in our own Solar System. They also show that a stellar encounter can reproduce the observed edge to the outer Solar System, marked by a sharp drop in the surface density of TNOs at a distance $r\sim45$ AU \citep{allen2001,morbidelli_kuiper_2003}. In addition, using both secular theory and numerical simulations, \cite{koncluster2020} investigated the effects of distant flybys on the inclination distribution of Kuiper belt objects. They found that the cold belt’s inclination can be increased by the observed amount, but the resulting distribution is incompatible with the data, thereby resulting in constraints on the dynamical history of the system.

Although the orbital distribution of the Sedna-like objects can help constrain dynamical processes in the outer Solar System, detection of these objects is relatively difficult as they reside in far distant orbits. In addition, observations of objects in our Solar System have been focused on low inclination orbits historically, and could be subject to observational bias. More recently, observational surveys have begun to explore regions with higher inclination. For example, the Dark Energy Survey recently detected 2015 BP$_{519}$, which has the highest inclination of any known TNO with a pericenter distance of greater than $35$AU \citep[]{becker_discovery_2018}. Future observations, including those conducted by the Large Synoptic Survey Telescope, are expected to dramatically increase the number of detected highly inclined Sedna-like objects \citep[]{trilling_detectability_2018}. These observations can be compared to theoretical models to constrain the architectural and formation history of our Solar System, such as stellar encounters that our Solar System could have experienced in its past.

In order to help interpret the aforementioned observational results, this paper characterizes the inclination distribution of Sedna-like objects under the influence of the passing stars. The resulting inclination distribution of Sedna-like objects depends sensitively on the flyby parameters. As a result, we explore a large parameter space for the stellar flybys, and obtain both analytical and empirical expressions that describe how the inclination distribution depends on the input parameters. Understanding the unique signatures in the inclination distribution could help one determine the origin of the objects in the outer Solar System with highly inclined orbits, as well as constrain the properties of the Solar System birth cluster.

This paper is organized as follows: Section \ref{sec:general} presents the set up and defines the parameter space of the stellar flybys studied herein, as well as presents our general results. Section \ref{sec:hr} focuses on the hierarchical regime for stellar encounters and derives analytical expressions for inclination excitation due to a distant flyby star. Then, in Section \ref{sec:nonhr}, we analyze the qualitative signatures of stellar encounters in the non-hierarchical region, as well as an empirical relationship between stellar encounter parameters and the resulting highly inclined objects. We then discuss the implications of our results for the birth cluster of our Solar System. Section \ref{sec:conc} presents our conclusions, along with a discussion of their implications. We have uploaded data sets of the results of our simulations to https://doi.org/10.5281/zenodo.3960748 so that they may be available for further studies or for comparison with observations of outer Solar System objects in the future.
\section{Numerical Setup} \label{sec:general}

In order to simulate stellar encounters within a stellar cluster, we ran N-body simulations using the \textsc{Mercury} \citep[]{Chambers_Mercury} code with the Bulirsch-Stoer integrator. We started with a debris disk composed of 10,000 test particles in circular orbits around a central Solar-type star, with semi-major axes randomly distributed uniformly between 30 AU and 300 AU. The orbits of these test particles initially all have vanishing mutual inclination, i.e. an idealized thin disk \citep[]{pringlediscs} which all particles lie within the same plane. This is in order to measure the degree of departures produced by the flyby interactions. Each test particle begins at a random place in its orbit. This scenario can be reduced to a gravitational three-body problem between the Sun, the flyby star, and each test particle \citep[]{halldisc,kobayashi_et_al_01,musielak3body}. Throughout this study we assume that self gravity between Sedna-like objects is negligible, since the total mass is low. In addition, we assume that the effects of the inner giant planets on the inclinations of Sedna-like objects is small, as we study objects with a pericenter distance ($r_p$) greater than $50$AU (we confirm this assumption later in this section). Moreover, we focus on objects with a semi-major axis ($a_t$) less than $10^3$AU so we neglect the effects of the galactic tide. We then study the effects of a stellar encounter on the debris disk.


The initial distance between the flyby star and the Sun is chosen to be
$$ R_o = 10,000 \,{\rm AU} (M_s/\mathrm{M_\odot})\,,$$ 
where $M_s$ is the mass of the stellar perturber. In addition, we stop the simulations after the flyby star has passed the Sun and returned back to a distance of $R_o$. To confirm that this $R_o$ value is large enough, we carried out simulations with an initial and final distance of $R_o = 10^5$AU (where $M_s = 1\mathrm{M_\odot}$ in this simulation). A KS test comparing the resulting cumulative inclination distribution from this simulation and the inclination distribution created by an encounter with the same parameters, but with $R_o = 10,000$AU, does not reject the null hypothesis at the 5\% significance level. Thus, we conclude that integrating the path of the star at distances greater than $(M_s/\mathrm{M_\odot})10,000$AU does not change our results. Once the flyby star has returned to a distance of $R_o$, we then measure the change in inclination of the test particles. For most of our analysis, we neglect test particles with a pericenter distance ($r_p$) of less than $50$AU post stellar encounter so that effects due to scattering and mean motion resonances with giant planets (which are not included in our simulations) can be neglected (recall that the 2:1 mean motion resonance with Neptune occurs at $a\sim48$ AU). 

\begin{figure*}[htb]
    \includegraphics[width=\textwidth]{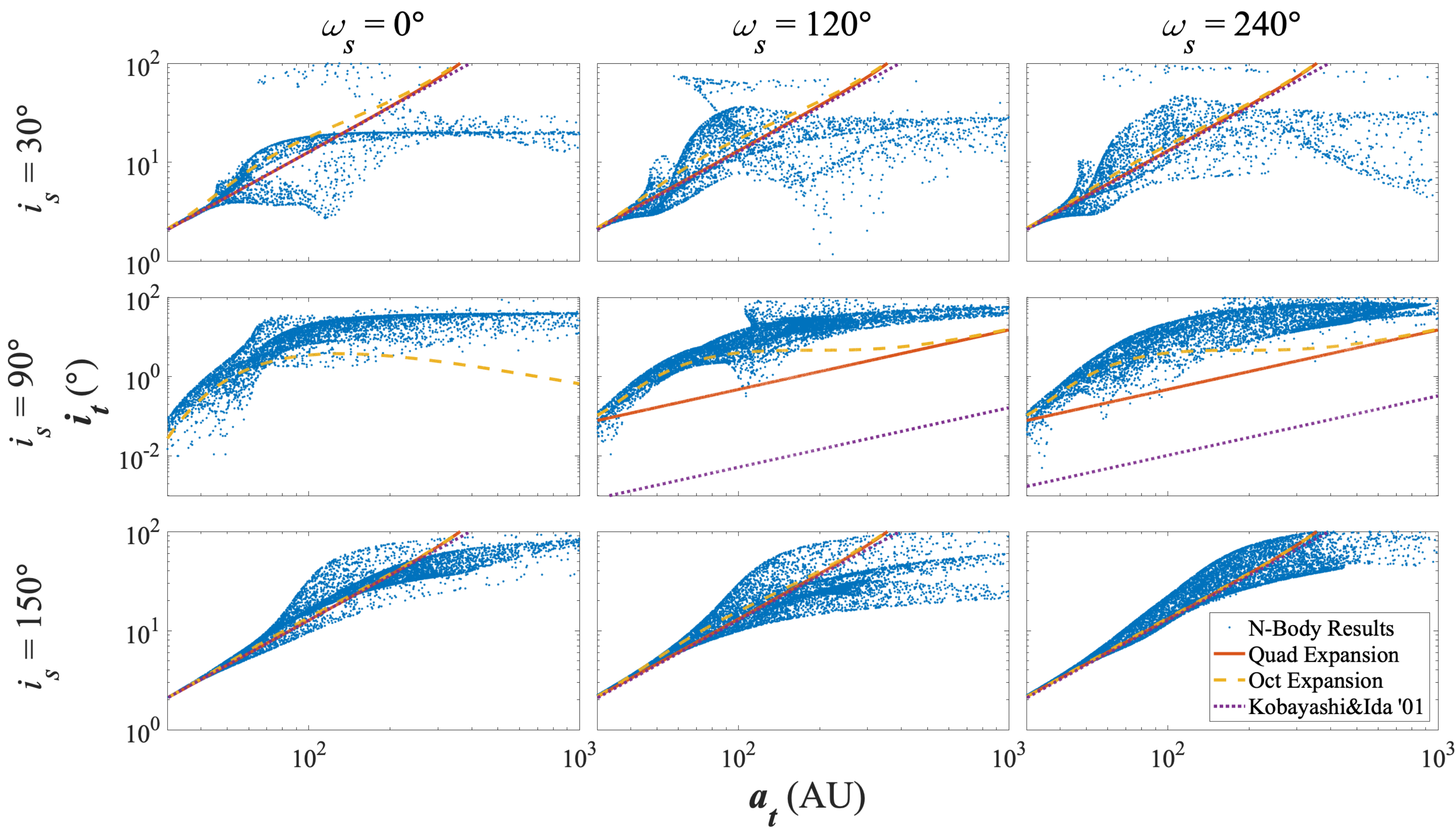}
    \caption{Inclination v.s. semi-major axis for test particles in a solar system debris disk post stellar encounter with pericenter distance of $R_p = 160$AU. Different columns correspond to different stellar incoming arguments of pericenter, and different rows correspond to different stellar inclinations. The solid red lines represent our analytical results at the quadrupole order, and the dashed yellow lines represent that at the octupole order (discussed in Section \ref{sec:hr}). The dotted purple lines represent the analytical expression obtained in \cite{kobayashi_et_al_01} to predict inclination excitation due to a stellar encounter. The expansion to octupole order is necessary to accurately predict the change in inclination of debris disk particle with highly inclined stellar flybys. Prograde encounters in general produce a higher variance in test particle inclinations and a small separated population at high inclination. Note that the quadrupole order predicts zero inclination excitation when $i_s=90^\circ$ and $\omega = 0^\circ$, so we do not include the quadrupole order results in the particular panel. \cite{kobayashi_et_al_01} predicts a negligible inclination excitation ($\lesssim10^{-14\circ}$) when $i_s=90^\circ$ and $\omega = 0^\circ$ and so we have omitted this result as well in the particular panel.}
    \label{fig:a_i_fit_23}
\end{figure*}

First, to illustrate the inclination excitation due to the geometry of a stellar flyby, we begin with a representative example. We start with a stellar encounter with a perturber mass $M_s = 1 \mathrm{M_\odot}$, and velocity at infinity $v_\infty$ = 1 km/s, typical speeds of young open stellar clusters \citep[]{li_cross-sections_2015}. The pericenter distance of the flyby star is taken to be $R_p = 160$AU, and we vary the incoming geometry of the stellar encounter. We consider five different values of stellar inclination $i_s$ in the range $30^\circ - 150^\circ$ in intervals of $30^\circ$, as well as the near coplanar encounter cases $i_s = 5^\circ, 175^\circ$. We also explore six values of stellar argument of pericenter $\omega_s$ in the range $0^\circ - 300^\circ$ in intervals of $60^\circ$.

A sample of the results with $\omega_s = 0^\circ, 120^\circ$ and $240^\circ$ and $i_s = 30^\circ$, $90^\circ$, and $150^\circ$ are shown in Figure \ref{fig:a_i_fit_23}. The blue dots represent the orbital parameters of disk particles after the stellar encounter. In this figure we keep particles with $r_p>30$AU (instead of $r_p>50$AU)  to show the accuracy of the analytical expression derived in Section \ref{sec:hr} at low semi-major axes. These simulations show that the inclination distribution of test particles have different signatures, with the most sensitive dependence determined by whether the stellar flyby was prograde or retrograde. In particular, prograde encounters tend to induce a greater variance in the inclination distribution, especially among test particles with larger semi-major axes. Another distinct signature of prograde encounters is that a sub-population of test particles is often excited to high inclination retrograde orbits and becomes distinguished from the main body of the distribution. This is consistent with the findings of \cite{breslau_pfalzner_retro_pro19} that shows prograde stellar flybys can create retrograde orbits.

It is important to note that an interesting symmetry exists between stellar encounters whose $\omega_s$ differ by $180^\circ$, while their other parameters are identical. Due to the azimuthal symmetry of our setup, the flyby stars in these cases share essentially identical paths; the only difference being whether the flyby star approaches the debris disk from above or below. This difference appears to be trivial as encounters of this type produce nearly identical inclination distributions. This finding halves the required parameter space needed to study the effects that stellar encounters have on inclination excitation in future simulations. For prograde stellar encounters with $\omega_s=0^\circ, 180^\circ$ the inclination excitation for particles with larger semi-major axis is nearly independent of the particle distances. We discuss the feature further in Section \ref{sec:bulk}.

\begin{figure}[htb]
\centering
\includegraphics[width=0.47\textwidth]{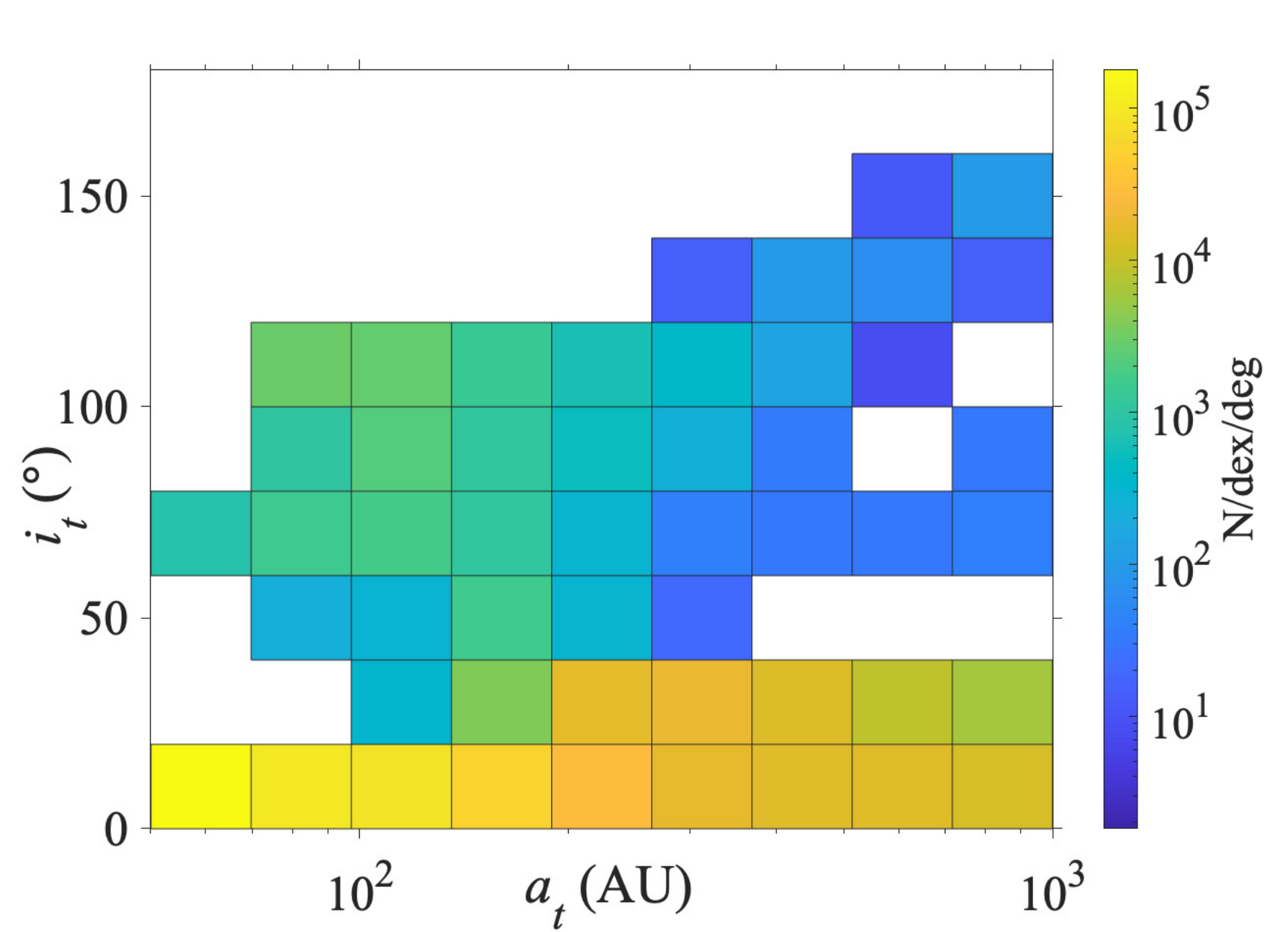}%
\caption{Density map of the test particles post stellar encounter corresponding to the upper left panel of Figure \ref{fig:a_i_fit_23}. The color corresponds to the number density within the bins. The high inclination population makes up a relatively small proportion of the entire distribution and the bulk of the particles is held at lower inclinations. 
}
\label{fig:dens}
\end{figure}

Figure \ref{fig:dens} shows the density map of the inclination distribution after the stellar encounter that corresponds to the the top left panel of Figure \ref{fig:a_i_fit_23}. This density map was constructed from an additional simulation of $10^6$ test particles. $\sim 5\times10^5$ particles are ejected from the system due to the stellar encounter and a further $\sim 2.5\times10^5$ particles are removed as their pericenter distances ($r_p$) are less than 50AU post encounter. We can see that the high inclination population created by prograde encounters only makes up a small fraction of the total distribution and that the bulk of the distribution is confined to lower inclinations. 

\begin{figure}[htb]
\centering
\includegraphics[width=0.47\textwidth]{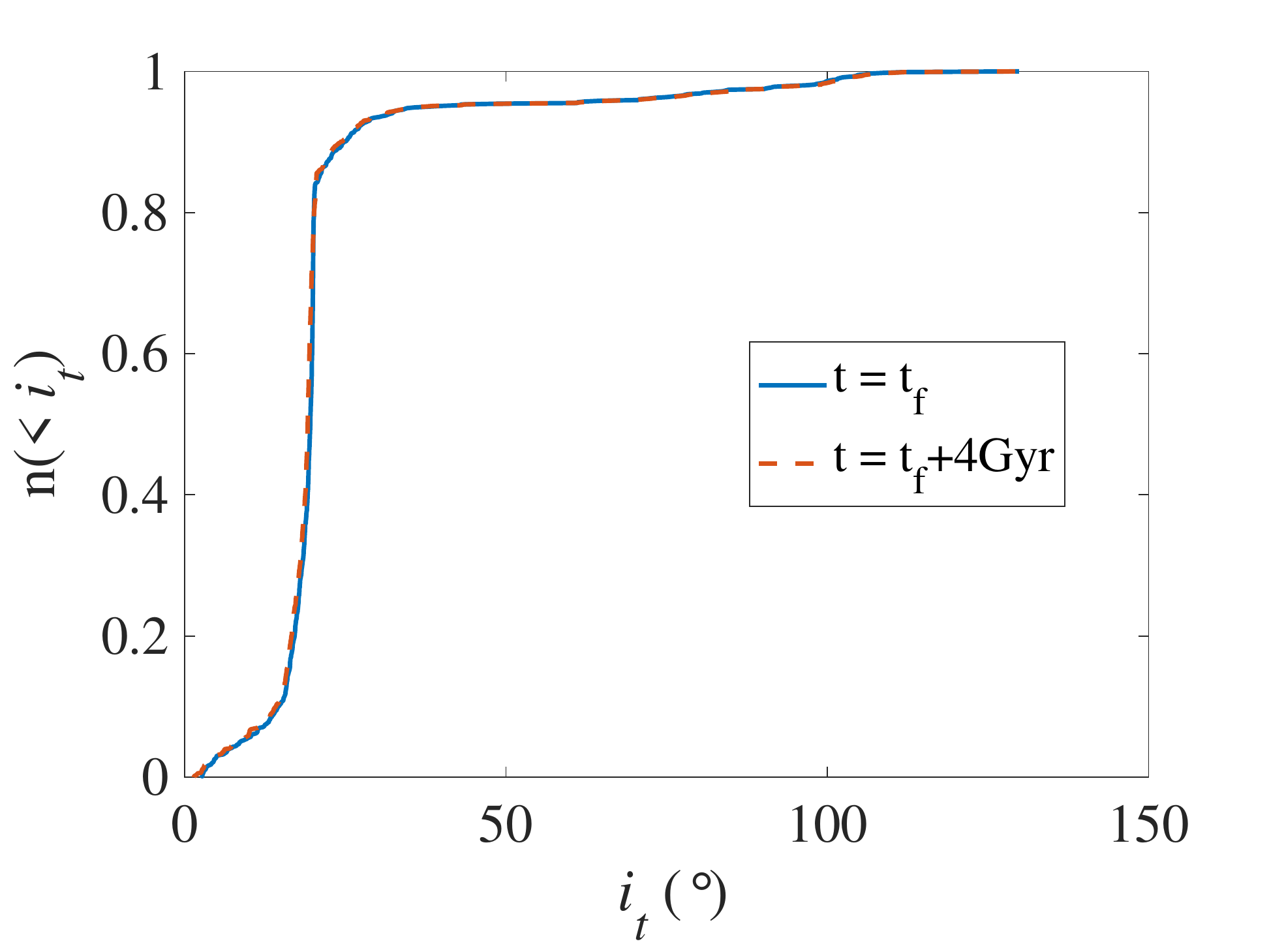}\\%
\includegraphics[width=0.47\textwidth]{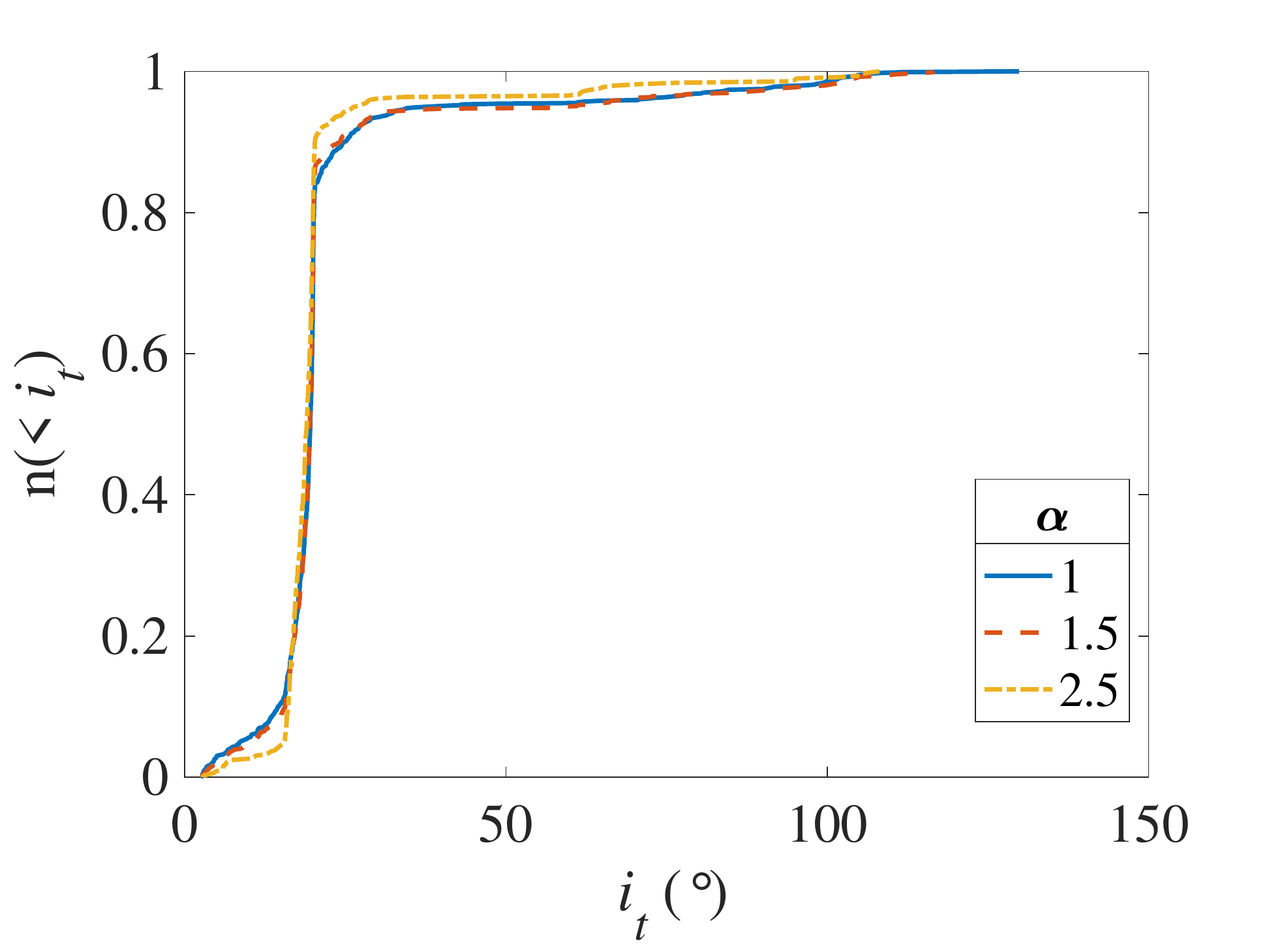}%
\caption{The {\bf top panel} shows the cumulative inclination distributions of particles immediately post stellar encounter (solid blue line) and that after a 4 Gyr evolution with the effects of the four giant planets (red dashed line). It shows that the inclination distribution remains relatively unchanged over the 4 Gyr. The {\bf bottom panel} shows the cumulative inclination distributions for debris disks with different initial surface densities ($\propto a_t^{-\alpha}$). The resulting inclination distribution does not change significantly for small variations in $\alpha$ ($\sim 1 - 2.5$).}
\label{fig:cdfs}
\end{figure}

\begin{figure}[htb]
\centering
    \includegraphics[width=0.47\textwidth]{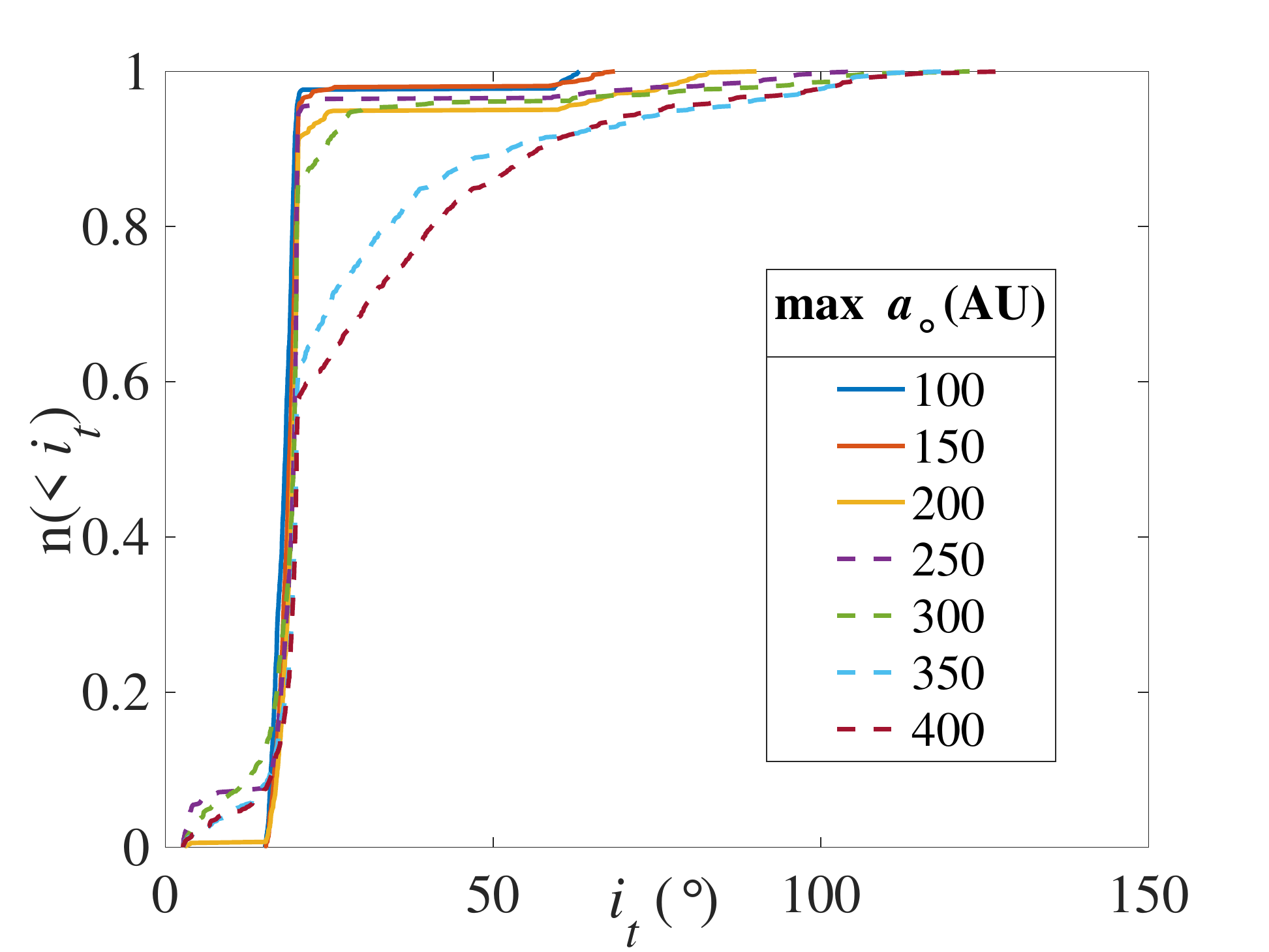}
    \caption{Cumulative inclination distributions for debris disks with different sizes ($a_{\circ,max}$). The different colors correspond to different initial outer edges of the disks. Debris disks with an outer edge of greater than 200AU are marked with dashed lines. Objects farther from the central star contribute to populations with moderate inclinations ($\sim 20-70^\circ$), which makes the inclination distribution smoother for large disks.}
    \label{fig:amax}
\end{figure}

To illustrate that the inclination distribution of the selected particles does not change significantly post the stellar encounter, we further evolved the perturbed debris disk for another 4 Gyr (comparable to the age of our Solar System). We consider an example with flyby star parameters: $M_s = 1 \mathrm{M_\odot}$, $v_\infty $= 1 km/s, $R_p = 160$AU, $i_s = 30^o$, and $\omega_s = 0^o$. These longer integrations include the effects of the J2 potentials resulting from Jupiter, Saturn, and Uranus in their present day orbits, as well as a Neptune as a point mass particle with its present day orbital parameters. After selecting particles with $r_p>50$AU we compare the resulting inclination distributions before and after this 4 Gyr evolution. Figure \ref{fig:cdfs} compares the cumulative inclination distributions. A KS test comparing these two distributions does not reject the null hypothesis at the 5\% significance level confirming their similarity. Although it is possible that the four giant planets migrated over the course of the lifetime of our solar system \citep[]{tsiganis_origin_2005, Levison_Disk}, they are likely to have been in tighter orbits in the past. This static model serves as conservative test and confirms that we can ignore the effects of the giant planets on debris disk inclination. 

For our simulations, we modeled our debris disk with a surface density profile $\Sigma \propto a_t^{-\alpha}$ where $a_t$ is the semi-major axis of test particles, and $\alpha=1$. However, the initial surface density of our Solar System's debris disk is unknown. Thus, we compared the inclination excitation of debris disks with different initial surface densities ($\alpha=$ 1, 1.5, and 2.5) to see if our results depend sensitively upon this parameter. We set the stellar encounter parameters to be $M_s = 1 \mathrm{M_\odot}$, $v_\infty$ = 1 km/s, $R_p = 160 \mathrm{AU}$, $i_s = 30^o$, and $\omega_s = 0^o$. The resulting cumulative inclination distributions are shown in Figure \ref{fig:cdfs}. The different colors in this plot denote the different initial surface densities of the disk. This figure shows that the inclination distributions after the  stellar encounters are all similar. A KS test comparing the distributions resulting from $\alpha$ = 1.5 and 2.5 to our default distribution of $\alpha=1$ does not reject the null hypothesis at the 5\% significance level which suggests that the initial surface density (with small changes of $\alpha$) of the debris disk does not significantly affect the resulting inclination distribution due to a stellar encounter.

We also ran a series of simulations where we changed the maximum radius of the debris disk ($a_{\circ,max}$) to see how the initial size of the debris disk changes the resulting inclination distribution post stellar encounter. The stellar parameters of the simulated encounters were from our standard case ($M_s = 1 \mathrm{M_\odot}$, $v_\infty$ = 1 km/s, $R_p = 160 \mathrm{AU}$, $i_s = 30^o$, and $\omega_s = 0^o$). We tested debris disk radii ranging from 100AU to 400AU in increments of 50AU. We then exclude debris disk particles with a pericenter distance of less than 50AU. Figure \ref{fig:amax} shows the cumulative inclination distributions resulting from these encounters. 

As we can see in Figure \ref{fig:amax}, the high inclination population only weakly depends on the size of the disk as long as the disk is around $\lesssim 100-300$ AU. At the low inclination end, debris disks with an initial radius of less than 200AU have few, if any, low inclination particles with a pericenter distance greater than 50AU post encounter. Most of the particles originating from these smaller debris disks are contained within moderate inclinations while a small percentage of particles exist at higher inclinations. On the other hand, debris disks with an initial radius of greater than 200AU contain low inclination populations. These larger debris disks also have high inclination populations that are more continuously connected to the bulk moderate inclination population. These signatures of low and high inclination populations can be compared to future observations of TNOs in our Solar System to constrain the initial size of our protoplanetary disk and can distinguish stellar encounters from other inclination excitation dynamical mechanisms. In Section \ref{sec:nonhr}, we study the dependence of the inclination distribution on the stellar flyby parameters.

\section{Hierarchical Regime} \label{sec:hr}
As illustrated in Figure \ref{fig:a_i_fit_23}, the inclination excitation can be divided into different regimes: 1) the low inclination excitation of test particles close-in to the central star and far away from the stellar flyby, 2) inclination excitation for particles with larger semi-major axis with greater variation in inclination, and 3) the inclination excitation of a small group of particles separated from the main distribution for the prograde stellar flyby. In this section, we start with the first regime, and consider the inclination excitation in the hierarchical limit, where the particles are far from the stellar flyby.

We use ${\bf r}=(x,y,z)$ to denote the position vector of the particle from the central star, and ${\bf R}=(\xi,\eta,\zeta)$ to denote that of the flyby star. The disk particles initially start in the $x-y$ plane. When the flyby star is far from the particles in the disk, one can expand the tidal force due to the flyby star in series of $r/R$  \citep[e.g.,][]{heggie_effect_1996,kobayashi_et_al_01}:
\begin{equation}
{\bf F}=\frac{GM_s}{R}\sum^{\infty}_{n=0} \nabla_{\bf r}\Big[\Big(\frac{r}{R}\Big)^n P_n\Big(\frac{{\bf r}\cdot{\bf R}}{rR}\Big)\Big] \,
\label{eq:forceexp}
\end{equation}
where $M_s$ is the mass of the flyby star and $P_n$ is the $n$th Legendre polynomial. Then, the change in the inclination can be estimated based on the change of the specific orbital angular momentum ($\mathbf{\delta h}$) due to the tidal torque. The change rate of the specific orbital angular momentum (torque acted by the incoming star) can be expressed as the following:
\begin{equation}
    \dot{\bf{h}} = \bf{r} \times \bf{F}
    \label{eq:torque}
\end{equation}


When the pericenter distance of the incoming star, $R_p$, is nearly comparable to the semi-major axes of the test particles in our simulations, we must use a non-secular approach for the calculation of $\mathbf{\delta h}$ (e.g., the ``The Exponential Regime'' where the eccentricity changes decay exponentially as a function of $R_p$ as discussed in \citealt{heggie_effect_1996}). We parameterize the particles' position as ${\bf r} = a_t ({\hat{\bf x}}\cos{(nt)}+\hat{\bf y}\sin{(nt)})$, where $a_t$ is the semi-major axis of the particle and ${\hat{\bf x}}$ and ${\hat{\bf y}}$ are the orthogonal unit vectors in the initial disk plane. This is different than the approach used in \cite{kobayashi_et_al_01} where they average the orbital phase of the particles. The azimuthal symmetry of the quadrupole potential makes it not a good approximation when the stellar encounter has a high inclination, and in particular it does not predict a change in inclination for stellar encounters with $i_s = 90^\circ$ if the argument of pericenter, $\omega_s=0^\circ$. Thus, we expand the perturbing acceleration $\vec{F}$ to the octupole order. This octupole expansion coupled with the non-secular treatment of the orbits of the particles is necessary to accurately predict the inclination excitation when $i_s\sim90^\circ$ as seen in the middle row of Figure \ref{fig:a_i_fit_23}.

Ignoring the more rapidly varying terms, $\bf \dot{h}$ to the first order in $nt$ can be expressed as the following:

\begin{align}
    {\bf \dot{h}}= \frac{3 a_t^2 G M_s}{8R^7}\Big(&(\zeta(4\eta R^2 + 10a_t\eta\xi\cos{(nt)} \nonumber \\
    &+ a_t(-4\zeta^2+11\eta^2+\xi^2)\sin{(nt)}), \nonumber \\ 
        &\zeta(a_t(4\zeta^2-\eta^2-11\xi^2)\cos{(nt)}\nonumber \\ 
        &-2\xi(2R^2+5a_t\eta\sin{(nt)})), \nonumber \\
        &+(-4\zeta^2+\eta^2+\xi^2)(\eta\cos{(nt)} \nonumber \\
        &-\xi\sin{(nt)})\Big)
        \label{eq:doth}
\end{align}

In the test particle approximation, the trajectory of the star is fixed. Thus, to estimate the total change of angular momentum ($\mathbf{ \delta h}$), we integrate $\dot{\bf h}$ over the course of the encounter along the path of the star. We adopted the ``method of steepest descent'' following \cite{heggie_effect_1996} to calculate the integral.


The variation in orbital inclination can be obtained as the following:
\begin{equation}
    \cos(\delta i) = \frac{({\bf h_o}+\mathbf{\delta h})\cdot {\bf h_o}}{||{\bf h_o}+\mathbf{\delta h}||\cdot||{\bf h_o}||} \,
    \label{eq:cosdi}
\end{equation}
where $\bf{h_o}$ is the initial orbital angular momentum. We expect ${\bf \delta h}$ to be small in the hierarchical limit, so we Taylor expand Equation \ref{eq:cosdi} around ${\bf \delta h}$ to obtain a simpler result. The inclination excitation at the lowest order of this expansion is:
\begin{equation}
    \cos(\delta i) \approx \frac{-\delta h_a^2 -\delta h_b^2}{2h_o^2} ,
    \label{eq:cosdisimp}
\end{equation}
where ${\bf \delta h} = (\delta h_a,\delta h_b,\delta h_c)$. 

The full expression for $\delta i$ to octupole order is lengthy so we leave it in its full form in Appendix \ref{sec:apA}. The quadrupole order of the expression is included below:

\begin{align}
\cos(\delta i) = &\frac{a_t^3M_s^2\sin^2(i_s)}{8a_s^3e_s^2(-1+e_s^2)^3m_\odot(m_\odot+M_s)} \label{eq:quad}\\ \nonumber
    &\times \Big[\Big(-3e_s\arccos(-1/e_s)\cos(i_s)\sin(\Omega_s) \\\nonumber &+\sqrt{1-1/e_s^2}((-1+e_s^2)\cos(\Omega_s)\sin(2\omega_s) \\\nonumber
    &+\cos(i_s)(-3e_s^2\\ \nonumber
    &+(-1+e_s^2)\cos(2\omega_s))\sin(\Omega_s))\Big)^2 \\\nonumber
    &+\Big(-3e_s\arccos(-1/e_s)\cos(i_s)\cos(\Omega_s) \nonumber \\ 
    &+\sqrt{1-1/e_s^2}\times(\cos(i_s)(-3e_s^2\nonumber \\ &+(-1+e_s^2)\cos(2\omega_s))\cos(\Omega_s) \nonumber \\
    &-(-1+e_s^2)\sin(2\omega_s)\sin(\Omega_s))\Big)^2\Big]\nonumber
\end{align}
where elements subscripted with ``t" refers to the debris disk object being perturbed, elements subscripted with an ``s" refers to the stellar perturber, and elements subscripted with an ``$\odot$" refers to the central star of our system.

We compare our analytical result to our N-body simulations in Figure \ref{fig:a_i_fit_23}. As shown in Figure \ref{fig:a_i_fit_23}, this analytical result is more accurate for particles with small semi-major axis. This is expected because we only considered the octupole expansion in the torque due to the incoming star. At larger distances, the spread in the inclination excitation is large and the detailed structure is rich. Thus, we do not expect that a single analytical formula could describe the inclination excitation at large distances.

\section{Non-Hierarchical Regime} \label{sec:nonhr}
In the non-hierarchical regime, the stellar flyby scatters the particles and excites their inclinations. The variance in the inclination excited is large (as shown in Figure \ref{fig:a_i_fit_23}), and it is difficult to obtain an analytical expression to describe the inclination increases in this regime. Thus, we describe qualitatively effects of the stellar flyby and the inclination excitation mechanism in section \ref{sec:bulk} and \ref{sec:highpop}. Then, we explore different stellar flyby parameters and obtain a simple empirical expression on the maximum inclination excitation in section \ref{sec:emp}. We apply our results to possible stellar flybys of our own Solar System in section \ref{sec:obs}.

\subsection{Bulk Inclination Variation} \label{sec:bulk}

\begin{figure*}[htb]
    \includegraphics[width=\textwidth]{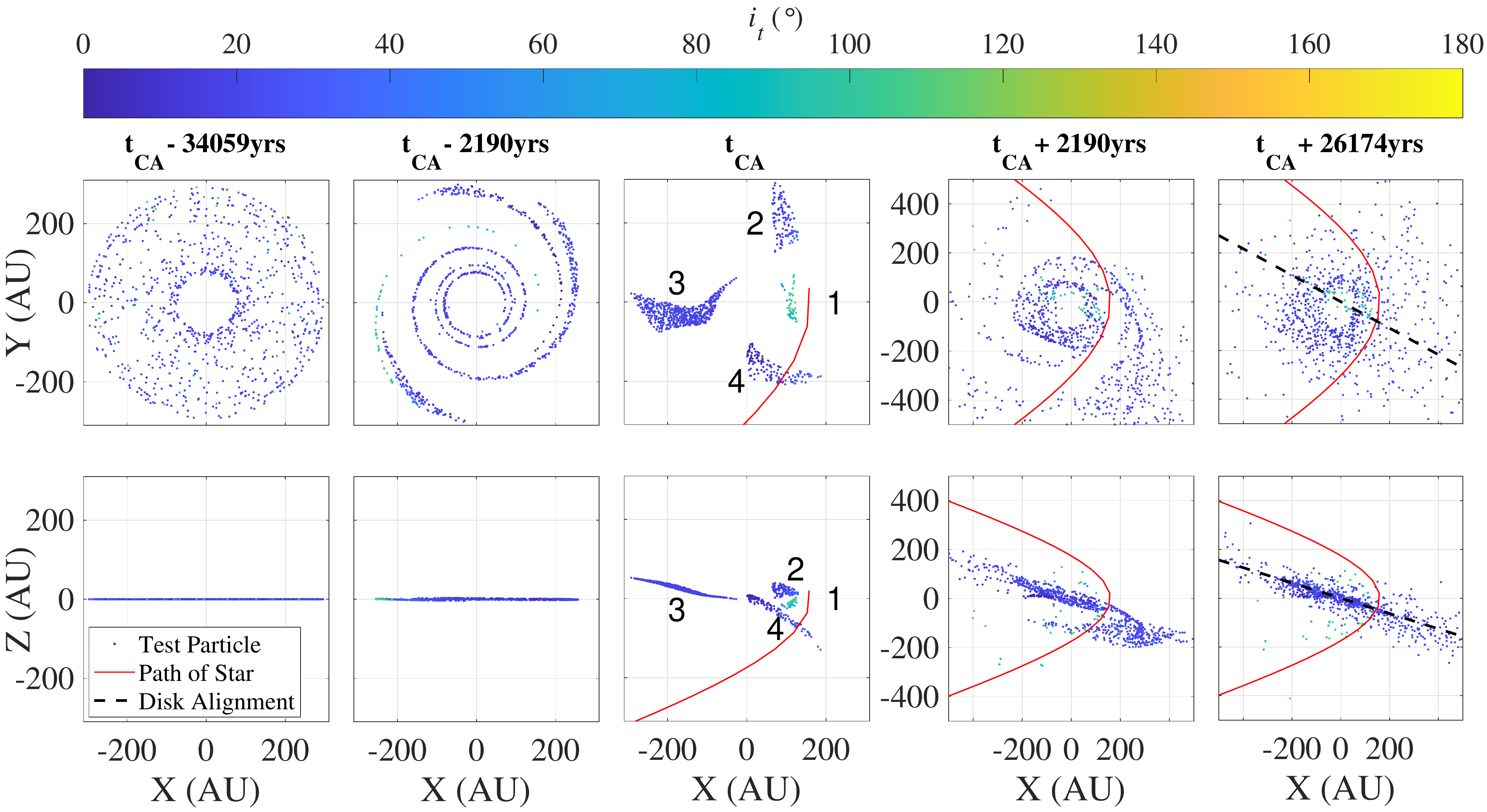}
    \caption{Snapshots of the survived test particles with a final $r_p>50$AU before, during, and after the stellar encounter. We denote the time of the closest approach by $t_{CA}$. The color of the dots represent the final inclination of the test particles at the end of the simulation. The red line shows the path that the star takes as it flies by the system, and the dashed black line shows the asymptotic exiting path of the perturbing star. The numbers in the middle panels denote four groups of test particles, where group $1$ corresponds to the highly inclined population. The overall tilt of the disk post encounter tends to align with the asymptotic exiting path of the star.}
    \label{fig:snapshots}
\end{figure*}

To investigate the detailed evolution of disk structure, we include in Figure \ref{fig:snapshots} the snapshots of the disk particles with a final $r_p>50$AU during a prograde stellar flyby. The parameters of the encounter are $M_s = 1 \mathrm{M_\odot}$, $v_\infty = 1 \mathrm{km/s}$, $R_p = 160$AU, $i_s = 30^o$, and $\omega_s = 0^o$. The upper row shows the positions of the particles in the $x-y$ plane, and the colors represent the final inclination of the particles. The particles start with azimuthal symmetry, as shown in the left panel when the star is farther away before the close encounter. As the star comes closer, spiral features can be produced due to the perturbation of the incoming star. The timescale of such spiral features is admittedly short ($\sim $kyrs), but this may suggest that some of the spiral features observed in debris disks by {\it ALMA} could be the result of stellar perturbation. 

The middle column shows the snapshot during the time of closest approach of the incoming star ($t_{CA}$). We can see that the survived particles with large final pericenter distances form four groups, where the group closest to the incoming star get its inclination maximally excited, corresponding to the high inclination populations shown in the upper row of Figure \ref{fig:a_i_fit_23}. This separation of groups is due to our selection of particles above pericenter of $50$ AU, and it is not seen in retrograde encounters. As the star moves past the closest approach, the spiral feature reappears, but as the star moves even farther, the spiral feature is gone and the disk appears azimuthally symmetric with particles scattered to larger separations from the host star. 

The bottom row shows the positions of the particles in the x-z plane. The particles initially lie within x-y plane as shown in the left panel. By the time of closest approach, long term perturbations caused by the advance of the flyby star have started to pull the particles out of their original alignment with the x-y plane as seen in the middle panel. As the star passes the system and continues on its journey, the orbits of most of the particles (those in groups labeled 2, 3, and 4) align with the asymptotic exiting path of the star. This behaviour can be seen in the bottom right panel.

The lower row of Figure \ref{fig:snapshots} shows that the disk seems to align with the asymptotic exiting path of the flyby star. The angle between the original orbital plane of the test particles and the asymptotic entering or exiting path of the flyby star can be obtained geometrically:
\begin{equation}
    \theta_s = \arccos{\sqrt{\cos^2{(\psi_s+\omega_s)}+\cos^2{i_s}\sin^2{(\psi_s+\omega_s)}}}
    \label{eq:theta_s}
\end{equation}
where $\omega_s$ and $i_s$ are the argument of pericenter and inclination of the flyby star respectively. $\psi_s = \pi-\arccos{(1/e_s)}$ is the scattering angle of the flyby star along the exiting path and $e_s$ is the flyby star's eccentricity. 

However, in exploring other stellar encounters with different geometries, we find that the disk only tends to align with the flyby star's path when the flyby star's pericenter lies within the original orbital plane of the debris disk. 
For near co-planar encounters the difference between $\theta_s$ and the average $i_t$ value is small, even for varying values of $\omega_s$. On the other hand, when the flyby star's trajectory is largely tilted from the disk plane, the particles cannot follow the path of the star. The disk is significantly warped and Equation \ref{eq:theta_s} is no longer a good approximation.

\subsection{Highly Inclined Particles} \label{sec:highpop}
Figures \ref{fig:a_i_fit_23} and \ref{fig:snapshots} show that prograde stellar encounters can produce a small population of highly inclined objects \citep{breslau_pfalzner_retro_pro19}, and that the final inclination of an object is closely related to its position at the time of closest approach. The population that is closest to the star at the time of closest approach (labeled ``1'' in Figure \ref{fig:snapshots}) can have its inclination excited to above $90^\circ$.

\begin{figure}[ht]
\centering
    \includegraphics[width=0.47\textwidth]{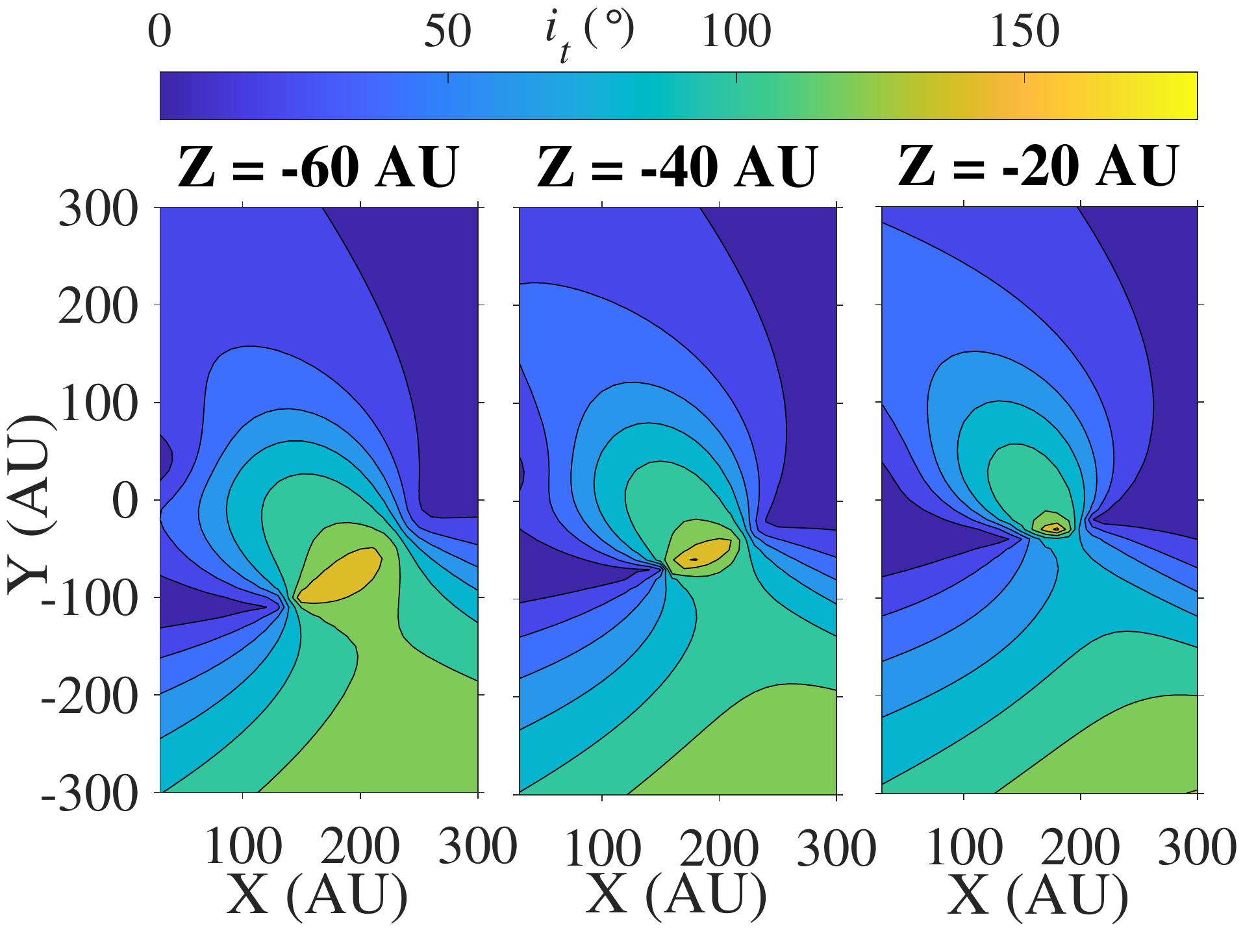}
    \caption{Predicted inclination excitation for test particles at different locations. High inclination excitation exists around $\sim20-60$AU below the original orbital plane of the disk and concentrates in a small region around $X\sim 100$AU and $Y\sim-80$AU, agreeing with the snapshots shown in Figure \ref{fig:snapshots}.}
    \label{fig:contour}
\end{figure}

To explain why this population experiences such extreme inclination excitation compared to the other groups of particles, we estimate the inclination excitation that stationary test particles would experience by calculating the torque exerted to the flyby star. Inclinations of the particles in group 1 changes rapidly near the closest approach of the flyby star over a small amount of time compared to the particles' orbital periods. Therefore, we assume the location of the particles are fixed for simplicity. We integrate the torque from the star onto the particles over the course of the stellar encounter to find the change in angular momentum of each particle. We then calculate the corresponding change in inclination according to Equation \ref{eq:cosdi}. The results are shown in Figure \ref{fig:contour}. This method tends to over-predict the final inclinations of particles, but is useful for illustrative purposes.

We considered the stellar encounter with $i_s = 30^\circ$ and $\omega_s = 0^\circ$ ($M_s = 1\mathrm{M_\odot}$, $v_\infty = 1 \mathrm{km/s}$, $R_p = 160 \mathrm{AU}$) in Figure \ref{fig:contour}. The color represents the predicted inclination post stellar encounter for particles at different locations. It shows that the high inclination excitation region lies below the original orbital plane of the test particles in approximately the same location of the group of particles labeled ``1" in Figure \ref{fig:snapshots} during the encounter. Long-term interaction with the star is responsible for perturbing the test particles from their original orbits so that they can reside in the high inclination excitation region at the beginning of the short-term interaction. 

This complication explains why retrograde encounters do not produce distinct high inclination populations. The long-term interaction prior to the close encounter does not effectively separate the disk of surviving particles (with $r_p>50$AU) into separate groups. Because this separation is absent, the resulting inclination distribution is smoother for retrograde encounters.

\subsection{Empirical Results} \label{sec:emp}
\begin{figure}[ht]
\centering
\includegraphics[width=0.43\textwidth]{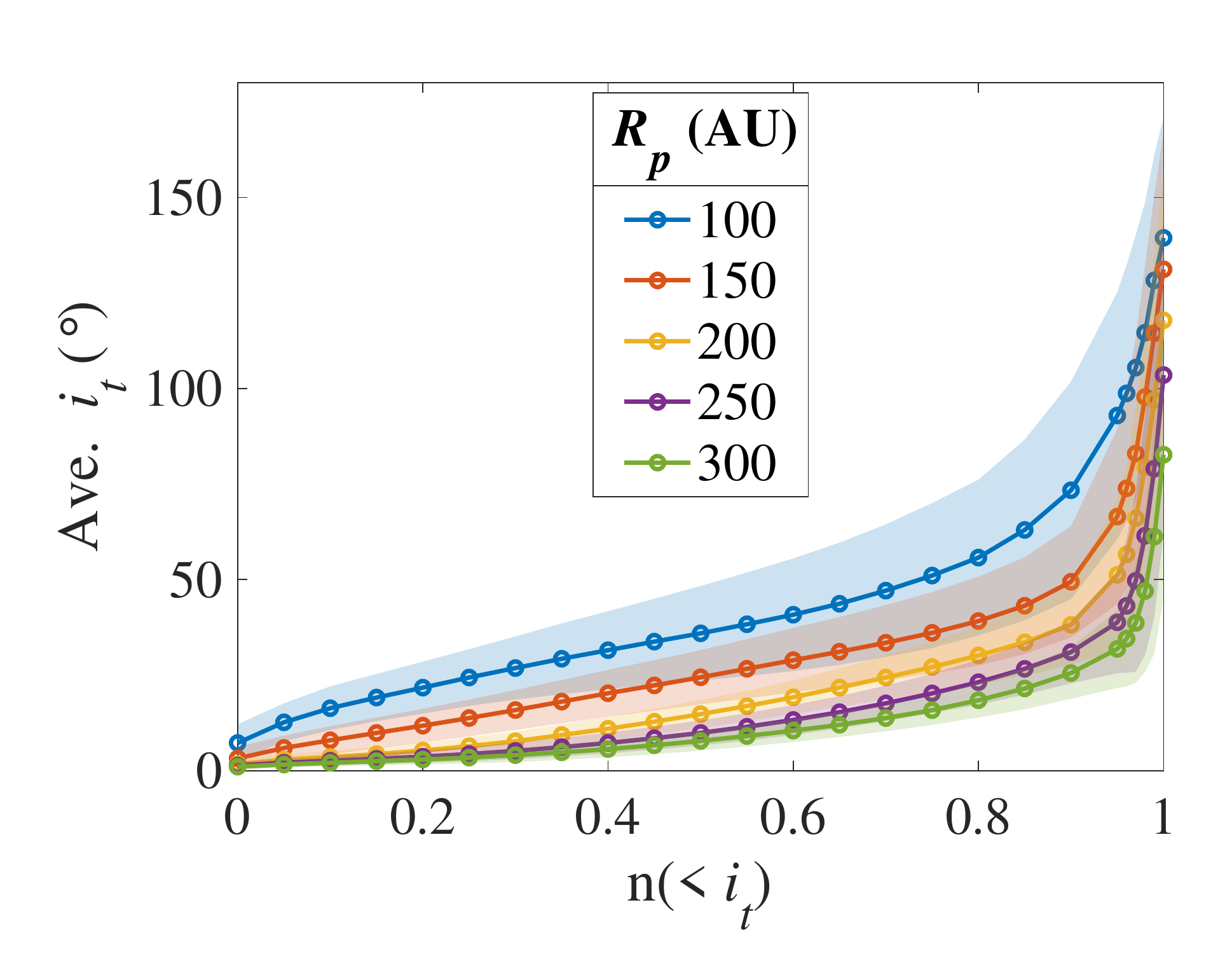}\\%
\caption{Weighted average of test particle inclinations post stellar encounter at different distribution percentiles. The inclinations are averaged over all $\omega_s$ and $i_s$, assuming the flyby is isotropic (weighted by $\sin(i_s)$). The shaded regions show the standard deviation around this weighted average. The different colors in each panel correspond to different $R_p$ of the perturbing star. We can see the general trend that encounters with smaller $R_p$ are more effective at exciting inclination. }
\label{fig:emp_rp1}
\end{figure}

A stellar pericenter distance ($R_p$) of 160AU helps recreate the observed edge of the Kuiper belt \citep[e.g.,][]{kenyon_stellar_2004}, and thus we use this value for most of our simulations. However, for a more complete exploration of encounter parameter space, we looked at the effects of varying $R_p$ on the resulting inclination distribution of the debris disk. The flyby star parameters for these tests are $M_s = 1 \mathrm{M_\odot}$, $v_\infty = 1 \mathrm{km/s}$. We choose $i_s$ between $10^\circ-170^\circ$ in intervals of $10^\circ$ (including the near co-planar encounter cases $i_s = 5^\circ, 175^\circ$, and twelve values of $\omega_s$ between $0^\circ-165^\circ$ in intervals of $15^\circ$. We vary $R_p$ between 50-300AU in increments of 10AU. The results are presented in Figure \ref{fig:emp_rp1}, which shows the inclination of debris disk particles averaged over all geometries of encounter, assuming the stellar encounters are isotropically distributed (weighted by $\sin(i_s)$ and assume $\omega_s$ is uniformly distributed). 

Figure \ref{fig:emp_rp1} shows a general trend that lower $R_p$ are more disruptive and can excite the inclination of particles to higher values. The shaded regions in Figure \ref{fig:emp_rp1} represent the standard deviation around the weighted average $i_t$, and illustrate that the geometry of encounter plays a significant role in the final inclination distribution of the debris disk. The specific geometry of encounter also becomes more important at lower $R_p$, as we can see that these encounters have greater deviation from the average due to the orientation of the perturbing star's path. 

\begin{figure}[ht]
\centering
\includegraphics[width=0.43\textwidth]{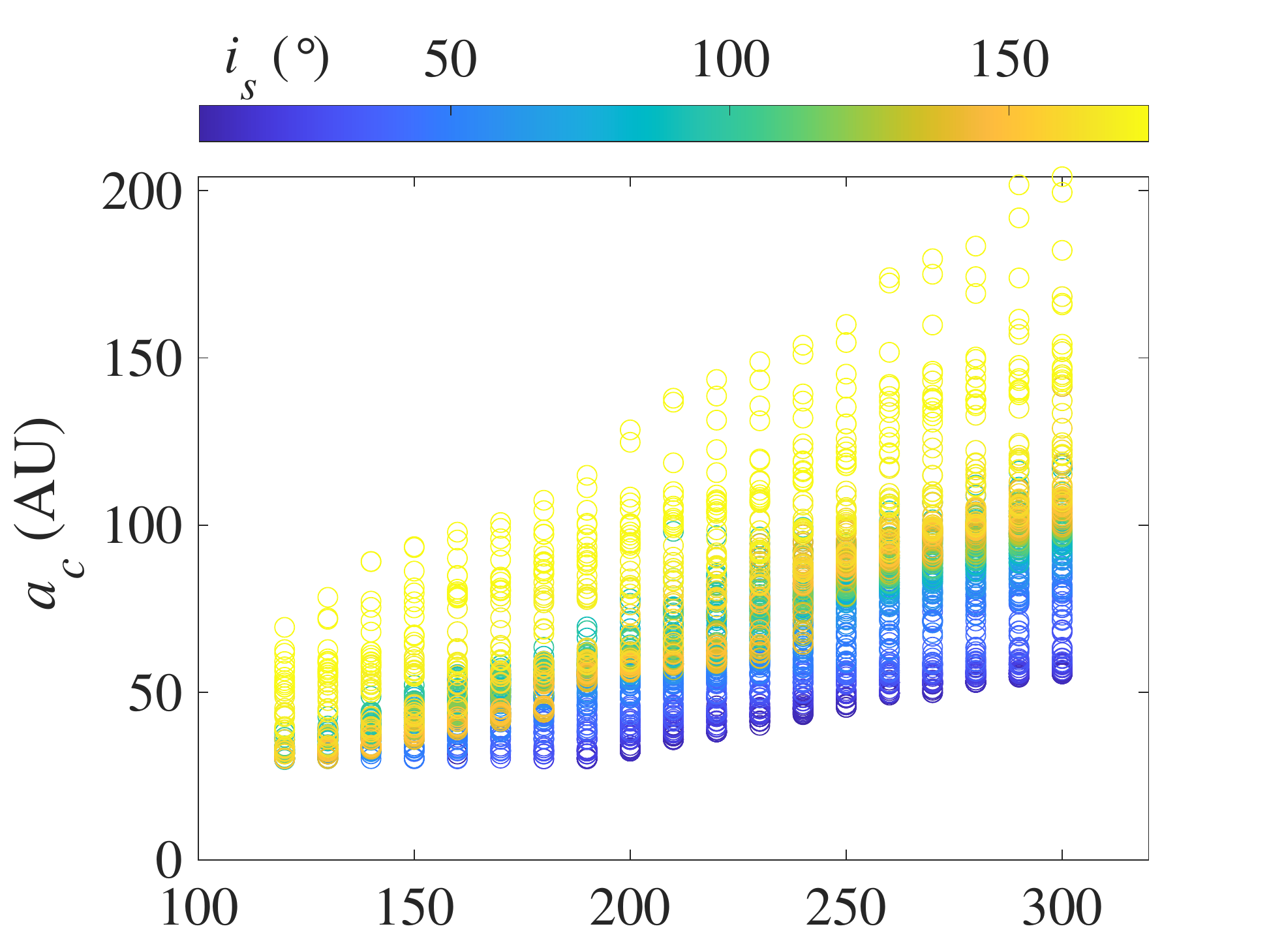}\\%
\includegraphics[width=0.43\textwidth]{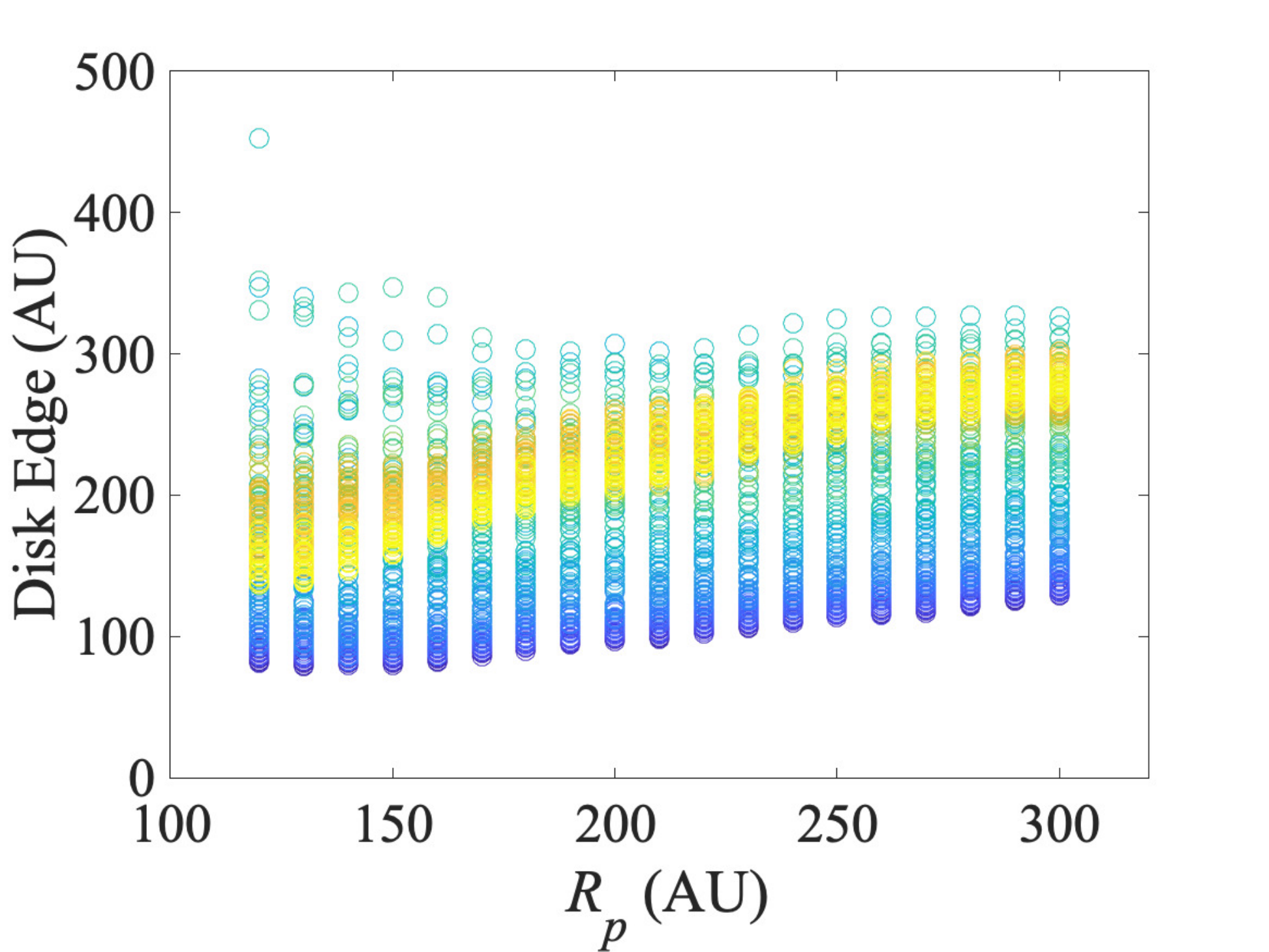}%
\caption{The {\bf top panel} shows the critical semi-major axis ($a_c$) of inclination distributions post encounter for different $R_p$. The colors denote the different $i_s$ of encounter. We can see that larger $R_p$ and $i_s$ lead to larger values of $a_c$. The {\bf bottom panel} shows the disk edge post encounter for different values of $R_p$. Generally, encounters with smaller $R_p$ lead to a smaller disk edge, and near coplanar encounters are most effective at shearing the edge of the disk.}
\label{fig:emp_rp2}
\end{figure}

Different $R_p$ leads to distinct inclination distributions, that can be characterized using critical semi-major axis ($a_c$) where the transition from the low variance hierarchical regime to the high variance non-hierarchical regime occurs. We define $a_c$ to be the semi-major axis at which the standard deviation of the inclination distribution exceeds 1 degree. The upper panel of Figure \ref{fig:emp_rp2} shows the values of ($a_c$) from stellar encounters with different $R_p$. We can see the general trend that encounters with larger $R_p$ lead to a higher $a_c$ in the disk post encounter. In other words, more particles fall within the hierarchical regime during encounters with larger $R_p$, and their inclination excitation are small. The inclination excitation of these particles can be predicted by our analytical work in Section \ref{sec:hr}. With lower $R_p$ ($<120$AU), an $a_c$ value does not exist for any geometry as the encounter is too disruptive. 

The geometry of encounter again plays a significant role in the location of $a_c$ as shown in the upper panel of Figure \ref{fig:emp_rp2}, where the colors represent different stellar flyby inclinations ($i_s$). Specifically, the lower inclination encounters are more disruptive and lead to lower values of $a_c$, while the less disruptive retrograde encounters tend to result in larger values of $a_c$. This can also be seen as an example in the top and bottom panels of Figure \ref{fig:a_i_fit_23} where the retrograde encounters have a smaller variance in $i_t$ at larger values of $a_t$. In the cases where $100\mathrm{AU}<R_p<150\mathrm{AU}$ most near coplanar prograde encounters are too disruptive for an $a_c$ value to exist. On the other hand, the dependence of $a_c$ on the stellar argument of pericenter ($\omega_s$) is weak. 

Stellar encounters have been used to explain the apparent edge of the Kuiper belt \citep{morbidelli_kuiper_2003,al2001,kenyon_stellar_2004,pfalzner_outer_2018}, marked by the sudden drop in surface density, and the edge of the disk post encounter also depends upon $R_p$ of the flyby star. The bottom panel of Figure \ref{fig:emp_rp2} shows the post encounter disk edge as a function of $R_p$ of the encounter. The location of the disk edge is identified to be where the surface density (in N/AU, where N is number of particles) of the disk falls to half of the peak value. We can see from this plot the general trend that larger values of $R_p$ lead to larger values of disk edge, though geometry of encounter still plays a significant role. Specifically, near coplanar prograde encounters tend to be the most effective at shearing the edge of the disk. Interestingly, different from $a_c$, counter orbiting ($i_s\sim180^\circ$) encounters produce smaller disk sizes comparing with the retrograde encounters with larger disk-flyby misalignment ($i_s\sim110^\circ$). Thus, observations of the inclination distribution as well as the surface density profile of debris disks can be used to help constrain $R_p$ and $i_s$ of past stellar encounters.

Next we vary the parameters of $M_s$ and $v_\infty$ in addition to $i_s$ and $\omega_s$. We choose five values of $M_s$ (0.1, 0.3, 1, 3, 10)$M_\odot$, eight values of $v_\infty$ (0.1, 0.3, 0.7, 1, 5, 10, 40, 100)km/s, $i_s$ between $10^\circ-170^\circ$ in intervals of $10^\circ$ (as well as the near co-planar encounter cases $i_s = 5^\circ, 175^\circ$, and twelve values of $\omega_s$ between $0^\circ-165^\circ$ in intervals of $15^\circ$. We again hold $R_p = 160$AU constant for these simulations.

\begin{figure*}[htb]
\centering
    \includegraphics[width=\textwidth]{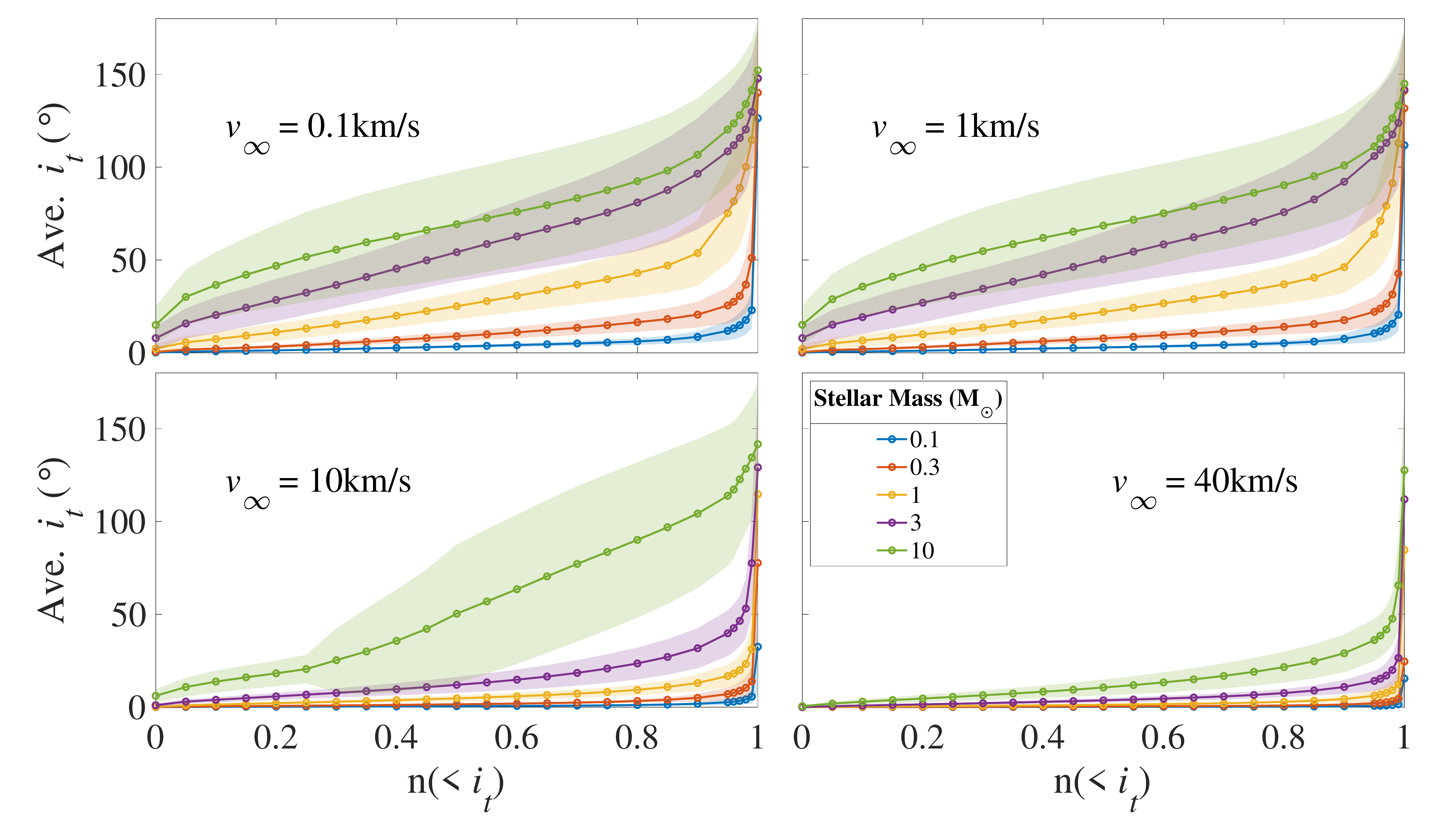}
    \caption{The weighted average of test particle inclinations post stellar encounter at different distribution percentiles. The inclinations are averaged over all $\omega_s$ and $i_s$, assuming the flyby is isotropic. The shaded regions show the standard deviation around this weighted average. The different colors in each panel correspond to different masses of the perturbing star. The different panels themselves correspond to different perturber velocities at infinity. From this figure we can see that larger mass and lower velocity stars can excite particles more effectively.}
    \label{fig:wave_fill}
\end{figure*}

Figure \ref{fig:wave_fill} shows the weighted average of debris disk particle inclinations post stellar encounter at different percentiles of the inclination distribution. The average inclinations for each $M_s$ and $v_\infty$ are calculated assuming the stellar encounters are isotropically distributed. In this figure we select four of the $v_\infty$ for illustrative purposes. We can see from this figure that larger mass perturbers are more effective at exciting inclinations of the debris disk particles than lower mass perturbers. In addition, perturbers with a lower $v_\infty$ are on average able to excite debris disk particles to higher inclinations than their faster counterparts. However, the average of the highest inclined debris disk particles tend to converge to similar inclinations at the highest percentile.

We also study the effect of varying the stellar perturber's mass and velocity on high inclination populations by investigating the maximum inclination after the stellar encounter. In order to remove outliers, we require that our maximally inclined particles have at least two other particles within $10^\circ$. This procedure does not remove the separated high inclination populations resulting from prograde stellar encounters (which may be separated from the main distribution by $>10^\circ$ in some cases) described earlier. We provide enough particles ($10^4$ particles in the disk) in our simulations for these separated populations to have well over 3 particles in all cases. We also require that our maximally inclined particles have semi-major axes less than or equal to $500$AU. Particles with $a_t>500$AU post encounter tend to have a chaotic dependence upon incoming stellar parameters and are thus less predictable. These particles are also much less likely to be observed due to their wide orbits.

\begin{figure}[htb]
    \includegraphics[width=0.47\textwidth]{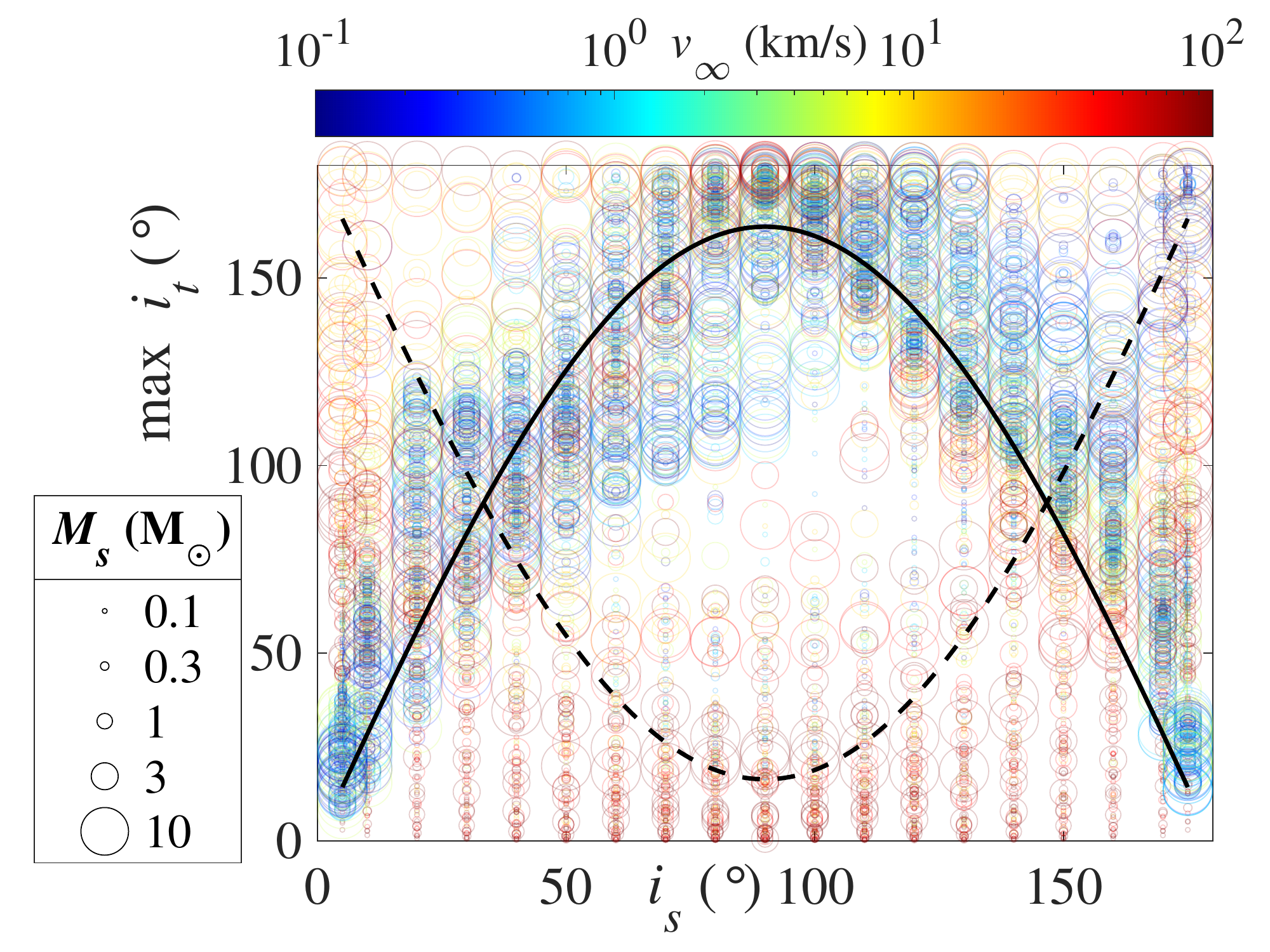}
    \caption{Maximum inclination of debris disk particles post stellar encounter versus stellar flyby inclination. The size corresponds to the mass of the stellar perturber and the color corresponds to the stellar velocity at infinity. The solid black line is $163.6327^\circ\sin(i_s)$ (fitted with $M_s \geq \mathrm{M_\odot}$ and $v_\infty \leq 5$km/s). The dashed black line is $180-163.6327^\circ\sin(i_s)$. More massive, slow perturbers tend to follow the positive $\sin(i_s)$ dependence, and the fast perturbers tend to follow the negative one.}
    \label{fig:enc_all}
\end{figure}

For prograde encounters, this maximally inclined particle belongs to the separated high inclination population. For retrograde encounters, the maximally inclined particle usually is drawn from the tail end of the distribution where $a_t$ is large post stellar encounter. 

Figure \ref{fig:enc_all} shows the maximum inclination of debris disk particles after the stellar encounter for every simulation. These maximum inclinations are plotted versus the inclination ($i_s$) of the fly-by star. From this figure we can see that most maximally inclined debris particles have one of two possible empirical relationships with $i_s$ as described below.

The first empirical relationship is shown in the solid black line of Figure \ref{fig:enc_all} and is given by $i_{t,max}=A\sin(i_s)$. The second empirical relationship is shown in the mirrored dashed black line of the same figure, given by $i_{t,max}=180^\circ - A\sin(i_s)$. As we can see from the color and size of the data points in this figure, which empirical relationship applies seems to depend mainly upon the perturber's mass and velocity. Specifically, flyby stars with more mass and lower velocities tend to favor the $i_{t,max}=A\sin(i_s)$ relationship, while flyby stars with less mass and higher velocities tend to favor the $i_{t,max}=180^\circ - A\sin(i_s)$ relationship. The amplitude $A$ is fit to the maximum inclinations of particles excited by stellar perturbers with $M_s \geq \mathrm{M_\odot}$ and $v_\infty \leq 5$km/s. The value $A=163.6327^\circ$ is calculated by minimizing the root-mean-square deviation between the empirical relationship and $i_{t,max}$.

We propose that there are two main processes through which a stellar perturber increases the inclination of debris particles. The first is long-term torquing of the orbit of the particles over the course of the encounter. In this case, encounters with an $i_s$ close to 90$^\circ$ have a mechanical advantage over near coplanar encounters in torquing the orbits due to the geometry of the encounter. This leads to the relationship $i_{t,max}=A\sin(i_s)$.

The second process is short-term scattering of the particles due to close proximity with the perturbing star. In the case where the perturbing star is too fast to effectively torque the orbits of the particles over a long period of time, the only way particles have their inclinations excited is by close short term scattering with the perturber. Therefore, when this process is dominant, near coplanar encounters will producer higher inclined particles as opposed to encounters with $i_s$ close to $90^\circ$ due to their proximity with the disk and leads to the $i_{t,max}=180-A\sin(i_s)$ relationship. The dependence on the stellar flyby argument of pericenter ($\omega_s$) is weak, except that Fast encounters ($v_\infty > 5$km/s) with $\omega_s\sim90^\circ$ are least effective at inclination excitation. This supports the picture that Fast encounters mainly excite particle inclination through short term scattering (which requires close proximity), as the pericenter of these encounters are furthest from the initial plane of the debris disk when $\omega_s\sim90^\circ$. ``Small" ($M_s < \mathrm{M_\odot}$) and ``Fast" encounters are capable of producing single objects with large inclinations but such close proximity scattering events are not effective in raising the average inclination of the entire disk being perturbed.

\subsection{Comparison to Observations} \label{sec:obs}
Here we compare our simulation results to observations of Sedna-like objects in our Solar System in order to constrain the parameters of possible stellar encounters that our Solar System experienced in its past. We take data from the {\sc IAU Minor Planet Center}\footnote{https://www.minorplanetcenter.net/} and select objects with $r_p>50$AU. This selection ensures that it is unlikely these objects had their orbits inclined via scattering with the planets. This selection also excludes most of the dominant regions of Mean Motion Resonances with Neptune. Currently this data set includes only nine objects, four of which have semi-major axes in the range 50--100AU while the rest have semi-major axes $a>100$AU up to $a\sim1000$AU. These objects have a wide range of eccentricities (0.11-0.93) and have inclinations ranging from $4.2^\circ$--$46.8^\circ$. The average inclination of these objects is $25.8^\circ$ with a standard deviation of $14.6^\circ$.


\begin{figure}[htb]
    \includegraphics[width=0.47\textwidth]{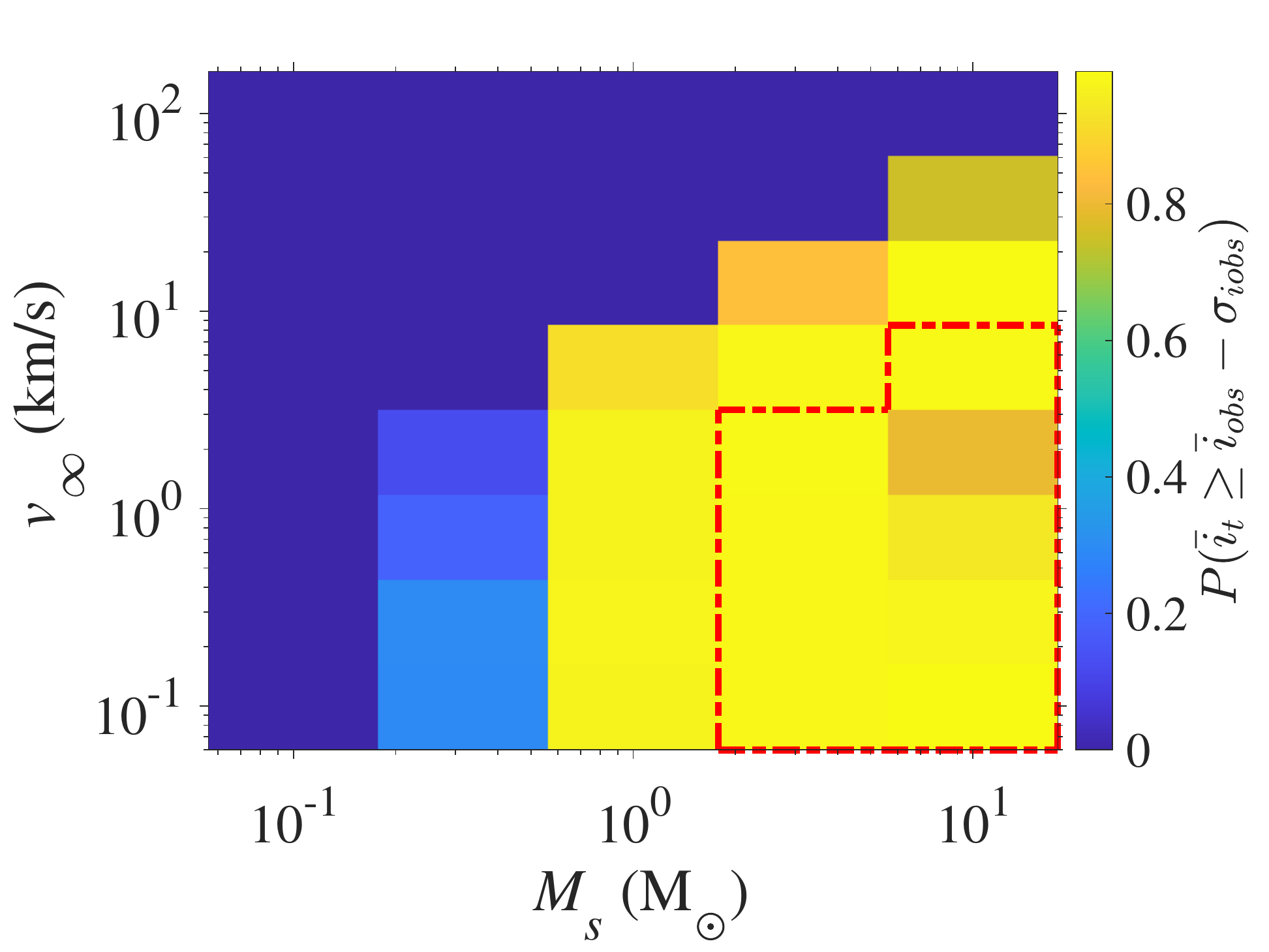}
    \caption{Probability of inclination excitation versus stellar flyby mass and velocity. The color represents the probability that the inclination excitation is higher than one standard deviation below the observed average ($\bar{i}_{obs}-\sigma_{iobs}$). The dashed red line denotes the region where inclination excitation mostly ($\gtrsim 80\%$) exceeds $\bar{i}_{obs}+\sigma_{iobs}$. It shows that stars with masses similar to that of the Sun with low velocity $\sim 1$km$/$s can lead to inclination excitation consistent with the observation (assuming $R_p=160$AU).}
    \label{fig:mv_prob}
\end{figure}

We first consider possible mass and velocity of the stellar flyby needed to produce particle inclinations consistent with the observations. Because observations of our Outer Solar System have been primarily focused on the objects near the ecliptic, we first use the observational result as a lower bound. We calculate the probability that the resulting average inclination is above one standard deviation lower than the observed average ($\bar{i_t} > \bar{i}_{obs} - \sigma_{iobs}$), while assuming that the distribution of stellar flybys is isotropic. Figure \ref{fig:mv_prob} shows this probability as a function of perturber mass ($M_s$) and perturber velocity ($v_\infty$). It shows that if a stellar perturber is not massive enough ($M_s\lesssim0.1\mathrm{M_\odot}$) or is too fast ($v_\infty\gtrsim40$km/s), then it is not sufficient to excite the average inclination. In this case, an additional dynamical mechanism would be required to explain the observed inclination distribution.

Next, we calculate the probability that the encounter will produce an average inclination below one $\sigma_{iobs}$ above the observed average ($\bar{i_t} < \bar{i}_{obs} + \sigma_{iobs}$). Only massive stars ($\gtrsim 2$M$_\odot$) with low incoming velocity ($\lesssim 1$km$/$s) could lead to inclinations higher than this observational constraint. Specifically, we mark the region where the probability is below $20\%$ in Figure \ref{fig:mv_prob} with a dashed red line. Stellar perturbers with these parameters are either so big or so slow that the average inclination exceeds what we currently observe. Please note that because of the aforementioned observational bias, this ceiling is likely to change with future observations of highly inclined objects.

In addition to the average inclination, we also compare our simulated inclination distributions to those of the observations using a KS test. Similar to the results using only the average inclination, the KS test shows that most of the encounters lead to distributions that are similar to the observations. Specifically, it does not reject the null hypothesis in $>90\%$ of encounters when $M_s=\mathrm{M_\odot}$ and $v_\infty\leq1\mathrm{km/s}$ at a $5\%$ significance level. Other encounters with higher stellar mass and velocity at infinity can favorably compare to observations as well, but specific geometries of encounter reject the null hypothesis when comparing to the observations in $\sim50\%$ of cases. The geometries that reject the null hypothesis in these cases tend to be nearly coplanar encounters ($i_s\leq10^\circ$ and $i_s\geq170^\circ$). Encounters with moderate to high stellar inclination ($30^\circ\leq i_s\leq130^\circ$) also reject the null hypothesis if their stellar argument of pericenter is below $60^\circ$, is between the range $135^\circ\leq\omega_s\leq240^\circ$, or is greater than $315^\circ$. We stress that relatively few observational data points exist for the types of objects we are interested in, so that it is difficult to do rigorous statistical comparisons. In the absence of sufficient data, our aim here is to give a rough estimate and to provide a framework that can be used when more observations become available. 

Combining both the inclination and eccentricity distributions can provide tighter constraints on the parameters of the stellar flybys \citep{pfalzner_outer_2018}. Toward this end, we have conducted two-dimensional KS tests that compare the inclination and eccentricity distributions of simulated and observed data simultaneously. We find that flyby stellar parameters that produce inclination distributions that compare favorably with observations ($M_s=\mathrm{M_\odot}$ and $v_\infty\leq1\mathrm{km/s}$) do not reject the null hypothesis in $\sim60\%$ of encounters when we compare inclination and eccentricity together. The specific geometries that reject the null hypothesis vary, but most tend to be near coplanar encounters ($i_s<30^\circ$ and $i_s>150^\circ$). Nearly perpendicular encounters ($80^\circ\leq i_s\leq130^\circ$) also tend to reject the null hypothesis when the stellar argument of pericenter is $\geq105^\circ$. This finding underscores the importance of the geometry of the encounter when attempting to reproduce multiple observed orbital parameter distributions simultaneously.

\begin{figure}[htb]
\centering
    \includegraphics[width=0.47\textwidth]{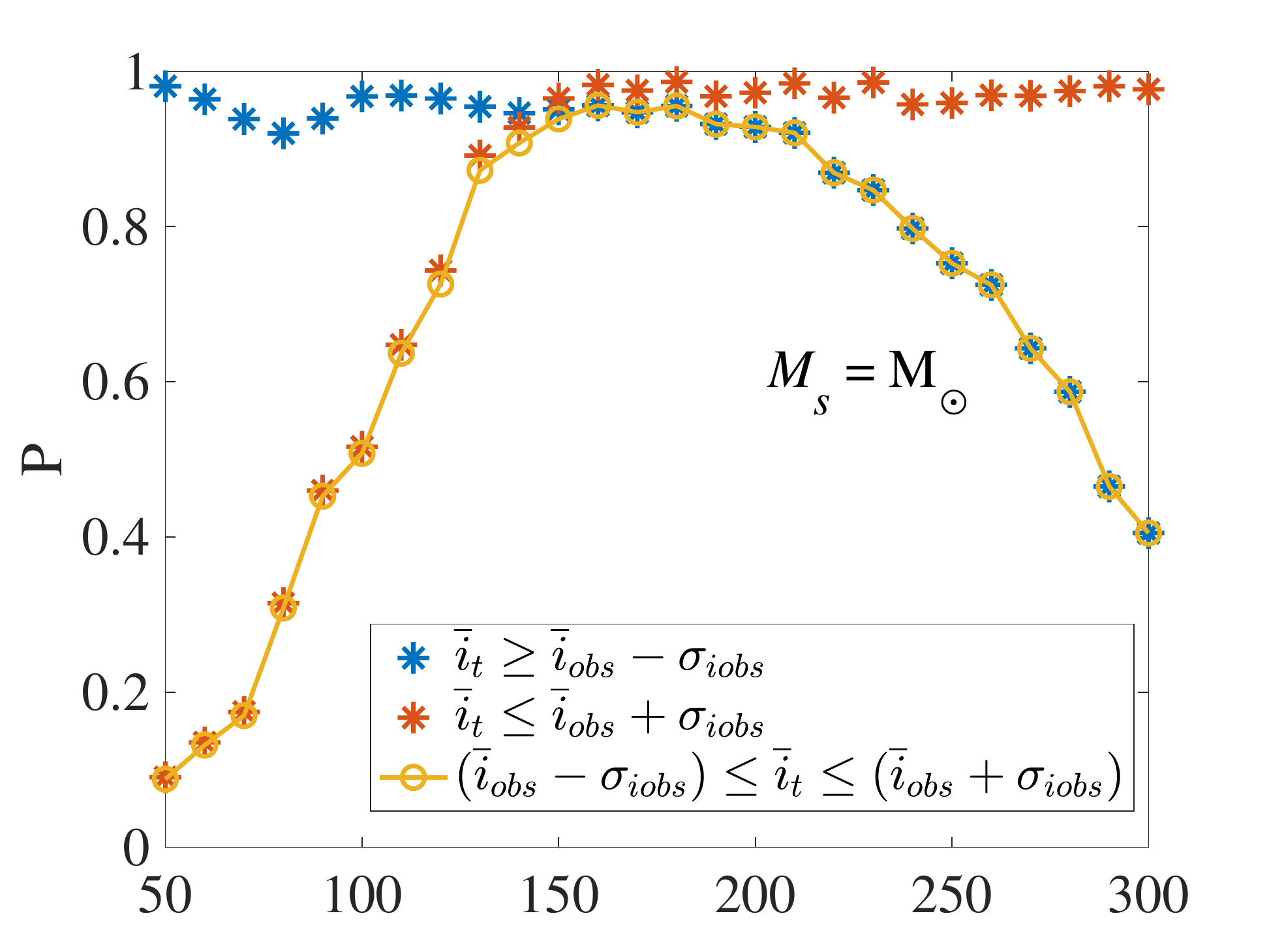}\\%
    \includegraphics[width=0.47\textwidth]{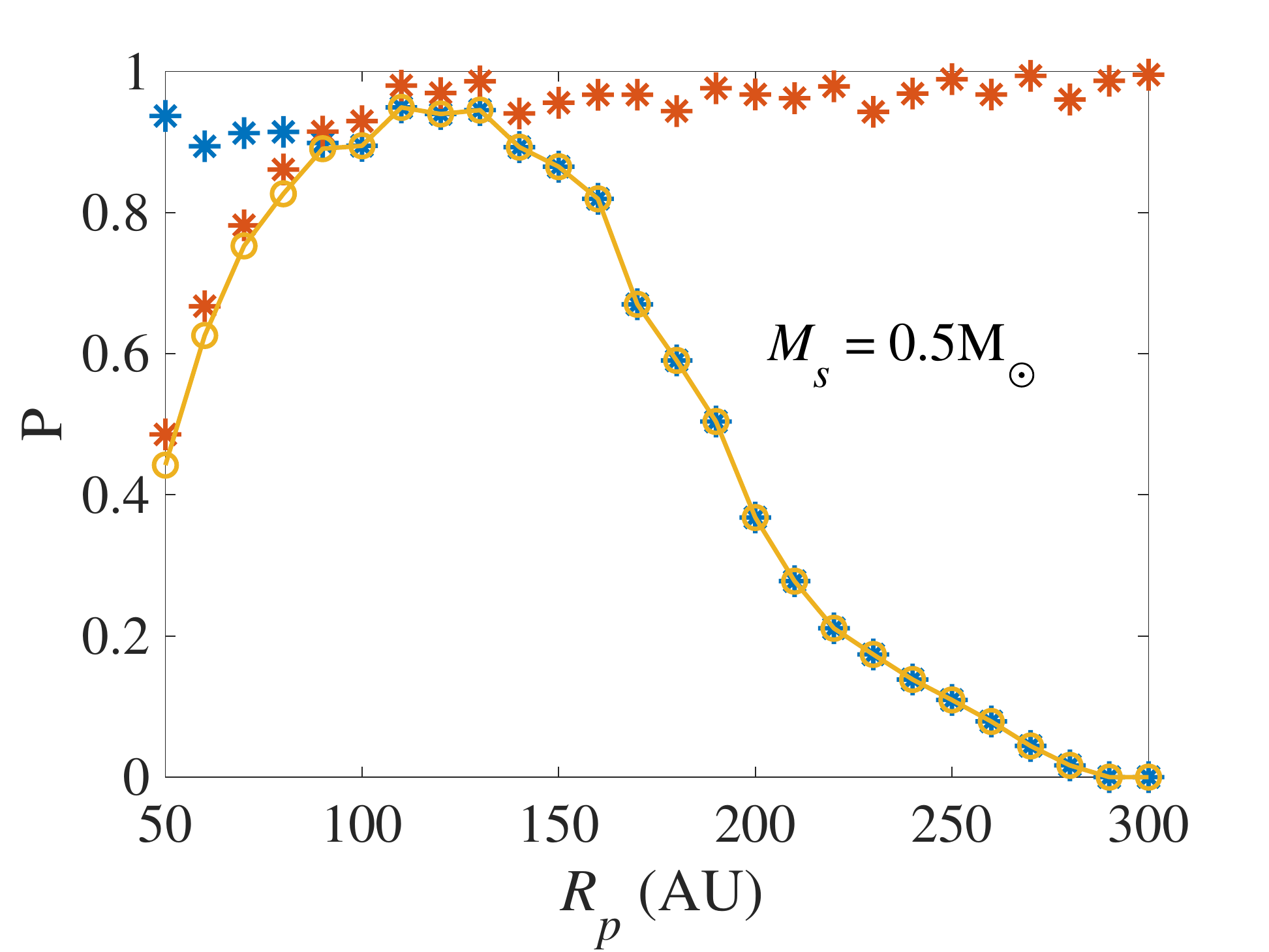}%
    \caption{The probability of inclination excitation versus stellar flyby pericenter distance. The blue stars show the probability of exciting inclination higher than one standard deviation below the observed average, and the red stars show the probability of excitation lower than one standard deviation above, $\bar{i}_{obs}+\sigma_{iobs}$. The yellow dots represent the probability of inclination excitation falling within one standard deviation of the observed average. The {\bf top panel} shows that a stellar flyby with mass $M_s=\mathrm{M_\odot}$ with pericenter distance around 160AU agrees well with the observed average inclination. The {\bf bottom panel} shows that a stellar flyby with $M_s=0.5\mathrm{M_\odot}$ with a shorter pericenter distance ($\sim120$AU) can also reproduce the observed average inclination}
    \label{fig:rp_prob}
\end{figure}

We next vary $R_p$ in our simulated encounters and hold constant $M_s=\mathrm{M_\odot}$ and $v_\infty=1$km/s. We can again calculate the average inclinations of distributions with these parameters and compare to the average inclination of Sedna-like objects in our Solar System with $r_p>50$AU. Figure \ref{fig:rp_prob} shows the probability of the inclination excitation in comparison with the observations. The blue stars show the probability of attaining inclination excitation higher than the observational constraint (one standard deviation below the average, $\bar{i}_{obs}-\sigma_{obs}$) and the red stars show the probability of falling below the observational constraint (one standard deviation above the average, $\bar{i}_{obs}+\sigma_{obs}$). These results imply that the inclination excitation agrees with the observations for $R_p$ between $\sim 150$AU and $\sim 200$ AU (assuming $M_s=\mathrm{M_\odot}$ and $v_\infty=1$km/s). The yellow dots show the probability that the simulated average inclination falls within one standard deviation of the observed average inclination. This probability peaks at around 160 AU. The average stellar mass in the cluster ($\sim0.5\mathrm{M_\odot}$) is lower than that of the Sun. For this smaller value $M_s=0.5\mathrm{M_\odot}$, we repeat this process and find that lower mass perturbers require a lower value of $R_p$ value, on average, to reproduce the observed inclinations. In this case, the most probable $R_p$ is $\sim120$AU. More generally, the required value of the pericenter of the flyby encounter roughly scales with mass according to $R_p \propto M_s^{0.5}$. 

Since $v_\infty$ is not universal in young clusters, we next find the most probable $R_p$ values for encounters with $M_s=\mathrm{M_\odot}$ at different values of $v_\infty$ using the same method. For encounters with $v_\infty=5$km/s, the most probable $R_p$ to reproduce the observed average inclination is between $\sim90$AU and $\sim110$AU. When $v_\infty=10$km/s, the most probable $R_p$ becomes $\sim70$AU. These types of faster encounters require the perturbing star to encroach further into the Solar System in order to excite the inclinations of Sedna-like objects to levels presently observed.

The most probable values of these flyby parameters are in rough agreement with previous studies. For example, \cite{kenyon_stellar_2004} finds that a solar-type star that passes by our Solar System on a prograde orbit at $R_p=160$AU can help explain the apparent edge (drop in  surface density) of the Kuiper Belt. They also find that such an encounter can also produce objects with Sedna-like orbits.  Work done by \cite{pfalzner_outer_2018} shows that fly-bys of stars on parabolic orbits with masses in the range
of $0.3 - 1.0\mathrm{M_\odot}$ at pericenter distances of between 50 and 150 AU at specific geometries can recreate many orbital features of the outer Solar System, including the dynamically distinct ``hot" and ``cold" populations of the Kuiper Belt and Sedna-like objects.

The results described above correspond to an interaction cross-section of $\sigma\sim 5\times10^5$AU$^2$ for producing the required orbital inclination. This value is roughly a factor of two larger than the disruption cross section estimated previously \citep{li_cross-sections_2015,Li2016}, which required that the giant planets were not strongly perturbed. In other words, flyby encounters can more readily overproduce orbital inclinations in the Kuiper Belt than perturb the orbits of giant planets. As a result, the corresponding constraint on the solar birth environment (from considerations of inclination excitation) is slightly more restrictive than constraints from orbital disruption of the giant planets (compare \citealt{li_cross-sections_2015} with \citealt{koncluster2020}). For a given disruption cross section $\sigma$, we require the optical depth for disruption to be less than unity, i.e., $n \sigma v\tau<1$, where $n$ is the number density of stars, $v$ is the typical speed, and $\tau$ is the time spent in the environment. Using $v=1$ km/s characteristic of intermediate-sized clusters in conjunction with the cross section of this paper, we find the limit $n\tau\lesssim8\times10^4$ Myr pc$^{-3}$.

From this result, the membership size $N$ of the Sun's birth cluster can be roughly estimated. We adopt the empirical relation for the lifetime of the cluster ($\tau = 2.3{\rm Myr} \Big(\frac{M_c}{\mathrm{M_\odot}}\Big)^{0.6}$), where we assume the mass of the cluster to be $M_c = 0.5{\rm M}_{\odot} N$ \citep{lamers_disruption_2005}. We find the average stellar number density assuming $N/R^2 = {\rm constant}$, where $R$ is the size of the cluster \citep{lada_embedded_2003, Adams06}. With an average stellar number density of $\sim 100-500$ pc$^{-3}$ ($N/R^2 = 450-550 {\rm pc}^{-2}$), we find that the upper limit of the cluster size is $\sim 10^{4}$. This result is in agreement with previous work, which also indicates that the membership size of the birth cluster obeys the constraint $N \lesssim 10^4$  \citep[e.g.,][]{al2001,Portegies_Zwart09,adams_birth_2010,Pfalzner13,pfalzner_cradles_2020}. If one also considers lower limits on the size of the birth cluster by requiring enrichment of short-lived radioactive nuclei and/or the production of Sedna-like bodies by scattering and/or the truncation of the solar nebula at 30 AU by external radiation fields, then the implied range for cluster membership becomes $N=10^3-10^4$ (see \citealt{adams_birth_2010} for further discussion).



\section{Conclusion} \label{sec:conc}

Objects in the outer realms of the Solar System are relics of the planetary formation processes and can provide important clues on the formation of the Solar System. In particular, the highly inclined Sedna-like objects cannot be explained by scattering with Neptune, and require some additional dynamical mechanism to explain their origin. To better understand the origin of the highly inclined objects, we have investigated the inclination excitation of debris disk bodies under the influence of close stellar encounters. We then used these results to constrain the properties of the Solar System birth cluster based on current observations of the inclination distribution of Sedna-like objects. 

Our results indicate that stellar encounters often leave unique signatures in the inclination distribution of the debris disks. In the hierarchical regime, where debris disk objects are relatively close to the central star and the flyby star acts as a distant perturber, the resulting inclination excitation is low. For this case we have derived an analytical expression that closely predicts the expected inclination excitation. Equation (\ref{eq:quad}) presents this result in the quadrupole limit, whereas equation (\ref{eq:oct}) presents the octupole terms. 

In the non-hierarchical limit, the encounters are more disruptive, and the resulting variance in the inclination distribution is large. Prograde encounters generally produce a larger scatter in the inclination distribution, and a small population ($\sim 5\%$) of highly inclined objects tends to be separated from the main body of the inclination distribution (specifically, for particles with $r_p>50$ AU). Based on a large ensemble of numerical simulations, we obtained the following general trends on the dependence of the disk properties: smaller debris disks ($a_{\circ,max}\lesssim 200$ AU) end up with few, if any, low inclination detached objects with $r_p>50$AU. Larger debris disks ($a_{\circ,max}\gtrsim 300$ AU) result in a smoother inclination distribution after the stellar encounter. On the other hand, the slope of the initial surface density of the debris disk (with power-law index $\alpha$ ranging between $-0.5$ and $-2.5$) does not change our results significantly. 

The inclination distribution also depends on the parameters of the incoming star. The pericenter distance $R_p$ of the stellar perturber has a significant effect on the resulting inclination distribution. Most importantly, $R_p$ greatly influences the critical semi-major axis ($a_c$) which marks the boundary between the low inclination dominated regime and the high inclination dominated regime (which has larger scatter) of the overall inclination distribution after the encounter. The geometry of the encounter also affects $a_c$ as the less disruptive retrograde flybys create a larger regime that is dominated by low inclination. In addition, we obtained an empirical relation for the maximum inclination excitation of the debris disk particles as a function of the stellar flyby inclination. In the regime where the perturbing star is massive and slowly moving ($M_s \geq \mathrm{M_\odot}$, $v_\infty \leq 5$km/s), the maximum inclination can be fit with a function of the form max$[i]\sim$ $163.6^\circ\sin(i_s)$. On the other hand, when the flyby star is lower in mass and moves faster, the maximum inclination can be fit by max$[i]\sim$ $180^\circ-163.6^\circ\sin(i_s)$.

Comparing our results to the current population of objects in our outer Solar System, we find that an encounter with stellar pericenter distance $R_p$ $\sim120$ AU can reproduce the observed average inclination. This finding assumes that the incoming star has a mass of $\sim 0.5$M$_\odot$, which corresponds to the average stellar mass from the expected initial mass function \citep{adams_theory_1996, chabrier_galactic_2003}. This level of disruption implies a constraint on the solar birth cluster of the form $n\tau\lesssim8\times10^4$ Myr pc$^{-3}$, where $n$ is the number density of stars and the $\tau$ is the residence time.  We note that constraints based on the inclination distribution of TNOs are slightly more restrictive than those found previously due to disruption of the orbits of the giant planets. (e.g., see \citealt{li_cross-sections_2015}). 


This constraint is also in rough agreement with previous studies that suggest the membership size of the birth cluster is bounded by $N \lesssim 10^{4}$ and likely falls in the range $N$ = $10^3-10^4$ \citep{al2001,Portegies_Zwart09,adams_birth_2010,Pfalzner13,koncluster2020, pfalzner_cradles_2020}. Future observations of outer Solar System objects with large inclinations will improve this constraint.

\section*{Acknowledgement}
The authors thank the referee for helpful comments, which improved the quality of the paper. This work was supported in part by NASA grant 80NSSC20K0522 and 80NSSC20K0641.

\appendix
\label{sec:appendix}
\section{Inclination Excitation in the Hierarchical Regime}\label{sec:apA}
In this Appendix, we include the analytical expression for the inclination excitation with terms on the octupole level ($a_1/a_2^3$). We adopt the following notations: subscript 1 refers to the central star, subscript 2 refers to the flyby star, and subscript 3 refers to the debris disk object that is being perturbed. The exception to this notation is $R_p$, which refers to the distance of closest approach of the flyby star. We assume the debris disk objects are test particles with zero masses.


When Equation (\ref{eq:forceexp}) is expanded to the octupole order, the resulting inclination increase (expressed as the quantity $\cos{(\delta i)}$) is given by the expression
\begin{equation}
\begin{split}
    \cos{\delta i}_{oct}=& \frac{(1+e_2)^4(m_1+m_2)^{5/2}m_2^2\pi R_p^{5/2}\sin^2(i_2)}{25,600 a_3 e_2^6(-1+e_2^2)m_1} \\
    & \times\Bigg[-a_3^{-3/2}(m_1+m_2+m_3)^{-7/2} \\
    & \times4\exp\Big(2\sqrt{\frac{m_1+m_2}{m_1+m_2+m_3}}(R_p/a_3)^{3/2}\big(-\sqrt{-1+e_2^2}+\arccos(1/e_2)\big)(-1+e_2)^{-3/2}\Big) \\
    & \times (-1+e_2)^{3/2} \bigg(\Big(9-9\cos(2i_2)+10\cos(i_2-2\Omega_2)+3\cos\big(2(i_2-\Omega_2)\big) \\
    &+18\cos(2\Omega_2)+3\cos\big(2(i_2+\Omega_2)\big)+10\cos(i_2+2\Omega_2)\Big)\sin(3\omega_2) \\
    &+\bigg(24\cos(i_2)+5\big(3+\cos(2i_2)\big)\Big)\cos(3\omega_s)\sin(2\Omega_2)\bigg)^2 \\
    &-\Bigg(16\exp\Big(2\sqrt{\frac{m_1+m_2}{m_1+m_2+m_3}}(R_p/a_3)^{3/2}\big(-\sqrt{-1+e_2^2}+\arccos(1/e_2)\big)(-1+e_2)^{-3/2}\Big) \\
    &\times R_p^{7/2}\bigg(-6\sin^2(i_2)\sin(\omega_2)\sin(2\omega_2) \\
    &+\cos^3(\omega_2)\Big(\big(1-10\cos(i_2)-11\cos^2(i_2)\big)\cos^2(\Omega_2)+4\sin^2(i_2) \\
    &+11\sin^2(\Omega_2)-\cos^2(i_2)\sin^2(\Omega_2)-5\cos(i_2)\sin(2\Omega_2)\Big) \\
    &-5\sin^3(\omega_2)\big(4\cos(i_2)\cos(\Omega_2)\sin(\Omega_2)+2\sin^2(\Omega_2)+\cos^2(i_2)\sin(2\Omega_2)\big) \\
    &+5\cos^2(\omega_2)\sin(\omega_2)\Big(2\sin(\Omega_2)\big(2\cos(\Omega_2)+\sin(\Omega_2)\big) \\
    &+6\cos(i_2)\sin(2\Omega_2)+\cos^2(i_2)\big(-4\cos^2(\Omega_2)+\sin(2\Omega_2)\big)\Big) \\
    &+(1/4)\cos(\omega_2)\sin^2(\omega_2)\Big(-36-20\cos(i_2)+36\cos(2i_2)+30\cos(i_2-2\Omega_2) \\
    &+15\cos\big(2(i_2-\Omega_2)\big)+90\cos(2\Omega_2)+15\cos\big(2(i_2+\Omega_2)\big) \\
    &+30\cos(i_2+2\Omega_2)-30\sin(i_2-2\Omega_2)+30\sin(i_2+2\Omega_2)\Big)\bigg)^2\Bigg) \\
    &\times\Big(\big(a_3/(e_2-1)\big)^{-3/2}\big((m_1+m_2+m_3)R_p\big)^{-7/2}\Big)\Bigg]
\label{eq:oct}
\end{split}
\end{equation}

\bibliographystyle{hapj.bst}
\bibliography{msref.bib}

\end{document}